\documentclass[review]{elsarticle}

\journal{Water Resources Research}

\usepackage[utf8]{inputenc} % allow utf-8 input
\usepackage[T1]{fontenc}    % use 8-bit T1 fonts
\usepackage{hyperref}       % hyperlinks
\usepackage{url}            % simple URL typesetting
\usepackage{booktabs}       % professional-quality tables
\usepackage{amsfonts}       % blackboard math symbols
\usepackage{nicefrac}       % compact symbols for 1/2, etc.
\usepackage{microtype}      % microtypography
\usepackage{lipsum}
\usepackage{color,soul}     % for comment hl
\usepackage{amsmath}        % for align % Used for the boldsymbol.
\usepackage{graphicx}       % use graphics

\usepackage{subfig} % use multiple figures

\usepackage{multirow}

\usepackage{setspace}
\renewcommand{\vec}[1]{\boldsymbol{#1}}
\usepackage{color,soul}
\begin{document}

\begin{frontmatter}

\title{Physics-Informed Neural Network Method for Forward and Backward Advection-Dispersion Equations}

\author[PNNL]{QiZhi He\corref{mycorrespondingauthor}}
\cortext[mycorrespondingauthor]{Corresponding author}
\ead{qizhi.he@pnnl.gov}


\author[UIUC,PNNL]{Alexandre M. Tartakovsky\corref{mycorrespondingauthor}}
\ead{amt1998@illinois.edu}
\address[PNNL]{Physical and Computational Sciences Directorate, Pacific Northwest National Laboratory Richland, WA 99354}
\address[UIUC]{Department of Civil and Environmental Engineering, University of Illinois Urbana-Champaign, Urbana, IL 61801}

%\begin{keypoints}
%	\item Physics-informed neural network (PINN) method is proposed for forward and backward advection-dispersion equations.  
%	\item The PINN method has several advantages over some grid-based discretization methods for high P\'{e}clet number problems. 
%	\item The PINN method is accurate for the considered backward ADEs that otherwise must be treated as computationally expensive inverse problems.      
%\end{keypoints}

\begin{abstract}
	%Advection-dispersion equations (ADEs) are commonly used to describe transport phenomena in porous media. Even though mature discretization-based numerical methods for ADEs exist, some challenges still remain, especially when it comes to solving advection-dominated forward ADEs and diffusion-dominated backward ADEs. The latter problem usually arises in the source identification context and leads to numerically unstable grid-based solutions that require a form of regularization or should be treated as an inverse problem that is computationally more expensive because it requires solving the forward problem multiple times. 
	We propose a discretization-free approach based on the physics-informed neural network (PINN) method for solving coupled advection-dispersion and Darcy flow equations with space-dependent hydraulic conductivity. In this approach, the hydraulic conductivity, hydraulic head, and concentration fields are approximated with deep neural networks (DNNs). We assume that the conductivity field is given by its values on a grid, and we use these values to train the conductivity DNN. The head and concentration DNNs are trained by minimizing the residuals of the flow equation and ADE and using the initial and boundary conditions as additional constraints. The PINN method is applied to one- and two-dimensional forward advection dispersion equations (ADEs), where its performance for various P\'{e}clet numbers ($Pe$) is compared with the analytical and numerical solutions. We find that the PINN method is accurate with errors of less than 1\% and outperforms some conventional discretization-based methods for $Pe$ larger than 100. Next, we demonstrate that the PINN method remains accurate for the backward ADEs, with the relative errors in most cases staying under 5\% compared to the reference concentration field. Finally, we show that when available, the concentration measurements can be easily incorporated in the PINN method and significantly improve (by more than 50\% in the considered cases) the accuracy of the PINN solution of the backward ADE.
\end{abstract}

%\begin{keyword}
%\texttt{elsarticle.cls}\sep \LaTeX\sep Elsevier \sep template
%\MSC[2010] 00-01\sep  99-00
%\end{keyword}

\end{frontmatter}

%\linenumbers

% ===============================================================================
%\input{sections/outline}
\section{Introduction}
%
%% ---- ADE: Background -----
Advection-dispersion equations (ADEs) are an important class of partial differential equations (PDEs) that are used to describe transport phenomena in the fields of  hydrology~\cite{ingham2005fundamental} and hydrogeology~\cite{Patil2014}.
%ADEs are used to model the transportation of chemical or biological species surface and subsurface water systems. Moreover, the ADE are used to describe heat and mass transfer in industrial applications such as casting~\cite{prescott1996convection}. ADEs are also used in financial applications as a model of the asset prices in stock markets. 

We consider both forward and backward ADEs. In the former case, the initial condition at time $t=0$ as well as  boundary conditions are specified and the solution of ADE is found for later times. This is a well-posed problem with well-established numerical methods. Nevertheless, there are some challenges in numerically solving forward ADEs, mainly associated with the advection-dominated problems. In backward ADEs, the concentration is known at later (terminal) times and solutions are sought for earlier times. Backward ADEs arise in the source identification problems \cite{wilson1994backward,neupauer1999adjoint} and could lead to numerically unstable grid-based solutions that require a some form of regularization \cite{xiong2006two} or should be solved as an inverse problem that is computationally more expensive because it requires solving forward problems multiple times \cite{atmadja2001state}.

%% ---- Numerical methods -----
Numerical discretization-based methods, including the finite elements (FE) and finite differences (FD) methods, are commonly used for solving the Darcy flow equation and ADE. 
Discretization-based methods approximate the PDE solution with its values at a set of grid points distributed over the spatiotemporal domain. The discrete solution is obtained by discretizing the time and spatial derivatives of state variables.
It is worth noting that the space-dependent parameters such as hydraulic conductivity are usually given not as a continuous field but as a set of values at the grid points.   

%% ---- Difficulties in solving AD equation -----
The combination of an advection (first-order) term and a dispersion (second-order) term with an anisotropic dispersion tensor in ADEs present several challenges for numerical methods. 
For example, for advection-dominated transport (i.e., P\'{e}clet number $Pe >> 1$), the numerical solutions can develop  oscillations (over- or undershoot) or numerical dispersion ~\cite{pinder2013finite,huyakorn2012computational}. These two numerical issues are closely related, and a numerical scheme developed to reduce numerical dispersion generally causes oscillation, whereas the suppression of oscillation comes at the cost of increased numerical dispersion ~\cite{Wang1997}.

%% ---- Remedies ----- 
Errors in numerical methods for ADEs can be reduced using a smaller grid size, but this results in a higher computational cost. Several methods have been developed to reduce errors for a given grid size~\cite{E.Ewing2001}, including 
% * Upwind for FEM
 the upwinding methods~\cite{heinrich1977upwind,noorishad1982upstream,brooks1982streamline}. Based on the ``optimal'' upwinding concept, the streamline upwind Petrov-Galerkin (SUPG)~\cite{brooks1982streamline,Methods1986}, the Galerkin least squares~\cite{Hughes1989}, and the unusual stabilized FEM (USFEM)~\cite{franca1995bubble} methods have been developed to increase the stability of the standard polynomial FE methods by consistently adding diffusive terms to the variational forms of ADEs. 
% High-order for FDM
 Higher-order schemes have also been developed to improve the accuracy of FD methods for ADEs ~\cite{Noye1988,Ding2009,rigal1994high,Cecchi2005}.

%% ---- DNN----- 
In this work, we obtain solutions of ADEs and the Darcy equation using the so-called physics-informed neural network method (PINN) \cite{Raissi2019,Mao2020}. Recently, the PINN method was applied for estimating hydraulic conductivity $k(\vec{x})$, steady-state hydraulic head $h(\vec{x})$, and concentration $u(\vec{x})$ fields using sparse measurements of these fields \cite{he2020physics}. In the PINN method for parameter and state estimation, the deep neural networks $\hat{k}$, $\hat{h}$, and $\hat{u}$ are used to approximate the $\hat{k}$, $\hat{h}$, and $\hat{u}$ fields, respectively. These DNNs are trained using $k$, $h$, and $u$ measurements constrained by the steady-state ADE and the Darcy flow equation.
In many applications, system states (including hydraulic head and concentration) change over time and can be easily observed as time series at fixed locations. The assimilation of time-varying data in the PINN framework requires constraining the DNN training with time-dependent governing equations. In this paper, we demonstrate that the PINN method for training $\hat{k}$, $\hat{h}$, and $\hat{u}$ DNNs given a known $k(\vec{x})$ field and constrained by the Darcy equation and time-dependent ADE with known initial and boundary conditions is equivalent to solving the forward Darcy and ADE equations. We also show that when the concentration is known at terminal time $T$, the PINN method approximately recovers the backward solution of the ADE for time less than $T$. Finally, we show that the measurements of $u$ (when available) can be easily incorporated in the PINN method and significantly improve the accuracy of the PINN solution. The accuracy of the PINN method is investigated via comparison with analytical and numerical solutions.  

%While the above-mentioned numerical methods have been widely used for ADEs, they are still mesh-dependent. It is nontrivial to construct discretization and estiamte the parameters used in the stabilization schemes to reduce the numerical dispersion as well as unphysical oscillation. 
The PINN method approximates the solution of a PDE with a DNN. Unlike other approximates, e.g., Fourier series, a large DNN can represent any continuous and bounded functions~\cite{Hornik1989,cybenko1989approximation}. 
Different from discretization-based methods, time and space derivatives in the PINN method are evaluated   using automatic differentiation of the DNNs ~\cite{Baydin2015} that does not involve any numerical discretization. Then, the DNN's  coefficients are computed by minimizing the loss function that is the sum of the residuals of both the PDEs and initial and boundary conditions. 
The PINN method has been used to solve various PDEs, including the steady-state diffusion equations with space- and state-dependent diffusion coefficients \cite{Tartakovsky2020a}, continuity and momentum conservation equations describing the flow of Newtonian~\cite{Raissi2020,Mao2020} and non-Newtonian fluids~\cite{Kissas2020}, and solid mechanics~\cite{Samaniego2020,Rao2020}.
%, and additive manufacturing~\cite{Zhu2020}.
The main idea in the PINN method of using an artificial neural network as an approximation function to solve a PDE dates back to the 1990s, when shallow feed-forward neural networks were used to solve initial-boundary value problems ~\cite{Lee1990a,Lagaris1997,Lagaris2000a}. The new interest in using artificial neural networks for solving PDEs is mostly attributed to advances in automatic differentiation~\cite{Baydin2015},  optimization methods~\cite{Byrd1995,Kingma2014,Goodfellow2016}, and specialized hardware (such as a graphic processing unit (GPU)) that significantly simplified implementation of the DNN-based methods and enabled the training of large DNNs, which might be necessary to approximate the solutions of PDEs.

While some variants of the PINN method have been used for solving the advection-diffusion equations~\cite{Yadav2016,Pang2018,Khodayi-Mehr2019,Dwivedi2020}, the performance of these methods was unsatisfactory in comparison with the grid-based methods, especially for the advection-dominated transport, where solutions develop large gradients.
As far as we know, this is the first study where the PINN method is used for solving an ADE with the velocity variations and the anisotropic dispersion tensor, which present additional challenges for grid-based methods.

Our paper is organized as follows. Section \ref{sec:method} presents the PINN method for ADEs. In Sections \ref{sec:result} and \ref{sec:result_back}, we present the PINN solutions of forward and backward ADEs, respectively. Section \ref{sec:data_assimilation} describes how data can be assimilated in the PINN method to increase the accuracy of the ADE solution. Finally, the conclusions are provided in Section \ref{sec:conclusion}.

%The paper proceeds as follows. Section \ref{sec:method} presents the formulations of advection-dispersion equaitons, followed by a brief introduction of physics-informed neural networks. Section \ref{sec:result} explains the test problems along with their numerical simulaiton results. Finally the conclusion is provided in Section \ref{sec:conclusion}.

%%%%%%%%%%%%%%%%%%%%%%%%%%%%
%%% Methods %%%
%%%%%%%%%%%%%%%%%%%%%%%%%%%%
\section{Problem formulation}\label{sec:method}
%%% ---------------------------------
% Advection-dispersion equation
%%% ---------------------------------
%
\subsection{Advection-dispersion equation}
We assume that transport in porous media is described by the ADE:
%
%\begin{equation}\label{eq:PDE_non}
%\left\{
%\begin{array}{lll}
%\begin{split}
%& u_t + \mathcal{N}_x[u] = 0,		  \quad  (\vec{x},t) \in \Omega \times (0, T)\\
%& u(\vec{x},t) 			= g(\vec{x},t),   \quad (\vec{x},t) \in  \partial \Omega \times (0, T) \\
%& u (\vec{x},0)    = h(\vec{x}),	    \quad \vec{x} \in  \Omega
%\end{split}
%\end{array} \right.
%\end{equation}
%where $\mathcal{N}_x$ is a spatial differential operator, and $g$ and $h$ are functions describing the boundary and initial conditions, respectively. 
\begin{equation}\label{eq:ADE}
\left\{
\begin{array}{ll}
\begin{split}
u_t + \nabla \cdot (-\vec{D} \nabla u + \vec{v} u) & = s, \quad (\vec{x},t) \in \quad  \Omega \times (0, T)\\
u & = g_D,  \quad  (\vec{x},t) \in \quad  \partial \Omega_D \times (0, T) \\
-\vec{D} \nabla u \cdot\vec{n}  & = g_N,  \quad  (\vec{x},t) \in \quad  \partial \Omega_N \times (0, T) \\
u (\vec{x},t=0) & = u_0, \quad  (\vec{x},t) \in  \quad  \Omega
\end{split}
\end{array} \right.
\end{equation}
where $\Omega \in \mathbb{R}^d$ is the spatial domain with the boundary $\partial \Omega$, $d$ is the number of spatial dimensions, $t$ is time,  $u(\vec{x},t)$ is the concentration, $\partial \Omega_D$ and $\partial \Omega_N$ are the Dirichlet and Neumann boundaries, respectively, $\vec{v}(\vec{x})$ is the average pore velocity $[LT^{-1}]$, $s$ is the source term $[ML^{-3}T^{-1}]$,  $g_D(\vec{x},t)$ and $g_N(\vec{x},t)$ are prescribed concentration and mass flux at the Dirichlet and Neumann boundary conditions, respectively, and $u_0(\vec{x})$ is the initial condition. The dispersion coefficient $\vec{D}$ $[L^2T^{-1}]$ is given as
\begin{equation}\label{eq:coe_dispersion}
% https://wwwbrr.cr.usgs.gov/projects/GWC_coupled/phreeqc/html/final-22.html
\vec{D} = D_w \tau \vec{I} + \vec{\alpha} {||\vec{v}||}_2,
\end{equation}
where $D_w$ is the diffusion coefficient, $\tau$ is the tortuosity of the medium, $\vec{I}$ is the identity tensor, and $\vec{\alpha}$ is the dispersivity tensor with the principal components $\alpha_{L}$ and $\alpha_{T}$. 
 In Sections \ref{sec:1D_AD}--\ref{sec:grid_effect}, we study a special case of Eq (\ref{eq:ADE}), where  $\vec{v}$ is known and constant in space and time and $\vec{D} = \kappa \vec{I}$ with a constant $\kappa$.
In Section \ref{sec:2DSTOMP}, we consider Eq (\ref{eq:ADE}), with $\vec{v}$ given by the solution of the Darcy equation with space-varying conductivity. In this test, we assume that the flow is steady and the solution is diluted, such that the velocity $\vec{v}$ is independent of $u$ and $t$.

\subsection{PINN approximation of the ADE solutions}

Below, we formulate the PINN method for the ADE (\ref{eq:ADE}). More discussion on the PINN method can be found in \cite{Raissi2019,Lu2019,Raissi2020,Mao2020,Tartakovsky2020a,he2020physics}.
In the PINN method, the solution $u(\vec{z})$ is approximated with a DNN as 
\begin{equation}
u(\vec{z}) \approx \hat{u}(\vec{z},\theta)  = \vec{y}_{n_l+1} (\vec{y}_{n_l}(...(\vec{y}_2(\vec{z}))),
\end{equation}
where $\hat{u}$ is a DNN approximation of $u$, $\theta$ is the vector of weights, $\vec{z}=[x_1,...x_d,t]$ is the space-time coordinate vector, and 
\begin{equation}
\begin{split}
\vec{y}_2(\vec{z}) &= \sigma(\vec{W}_1 \vec{z} + \vec{b}_1) \\
\vec{y}_{i+1}(\vec{y}_{i}) &= \sigma(\vec{W}_i \vec{y}_{i} + \vec{b}_i), i = 1,...,n_l-1 \\
\vec{y}_{n_l+1} (\vec{y}_{n_l}) &= \vec{W}_{n_l} \vec{y}_{n_l} + \vec{b}_{n_l}.
\end{split}
\end{equation}
Here, $n_l$ denotes the number of hidden layers, $\sigma$ is the predefined activation function, and the vector $\theta$ is defined as:
\begin{equation}
\theta = \{\vec{W}_1^T, \vec{W}_2^T, ..., \vec{W}_{n_l}^T, \vec{b}_1^T, \vec{b}_2^T, ..., \vec{b}_{n_l}^T\}.
\end{equation}
  
Substituting $\hat{u}(\vec{z},\theta)$ into Eq (\ref{eq:ADE}) yields the residual DNN:
\begin{equation}\label{eq:residual_f}
	r_f (\vec{z},\theta) = \hat{u}_t(\vec{z},\theta) + \nabla \cdot (-\vec{D} \nabla \hat{u} (\vec{z},\theta) + {\vec{v}} \hat{u}(\vec{z},\theta)) - s, \quad  \vec{z}\in  \Omega \times (0, T). 
\end{equation}
When $\vec{v}$ is non-uniform, it is also approximated with a DNN that is trained separately from $\hat{u}(z,\theta)$ because of the assumption that $\vec{v}(x)$ is independent of $u(x,t)$, as described in Section \ref{sec:2DSTOMP}. 

Similarly, the residual DNNs corresponding to the BCs and IC are obtained by substituting $\hat{u}(\vec{z},\theta)$ into the boundary conditions in Eq (\ref{eq:ADE}) as 
\begin{equation}\label{eq:residual_bc}
\begin{split}
r_{{BC}_D} (\vec{z},\theta) & = \hat{u}(\vec{z},\theta) - g_D(\vec{z}),  \quad  \vec{z} \in  \partial \Omega_D \times (0, T)	\\
r_{{BC}_{N}} (\vec{z},\theta) & = \vec{n} \cdot \nabla \hat{u}(\vec{z},\theta) - g_N ((\vec{z}),  \quad \vec{z} \in  \partial \Omega_N \times (0, T)
\end{split}
\end{equation}
and 
\begin{equation}\label{eq:residual_ic}
r_{IC}(\vec{x},\theta) = \hat{u} (\vec{x},t=0) - u_0(\vec{x}), \quad   \vec{x} \in \Omega.
\end{equation}
The residual networks in Eqs. (\ref{eq:residual_f})--(\ref{eq:residual_ic}), can be easily evaluated by applying automatic differentiation~\cite{Baydin2015} to the DNN $\hat{\vec{u}}(\vec{x},t;\theta)$. Next, we define the loss function  
\begin{equation}\label{eq:loss_pinn}
\begin{split}
J(\theta) =  w_f J_f(\theta) + w_{BC} J_{BC}(\theta) + w_{IC} J_{IC} (\theta),
\end{split}
\end{equation}
%$\omega_{\gamma}, \gamma \in \{ f, BC, IC \}$
where
\begin{equation}\label{eq:loss_components}
\begin{split}
J_f(\theta) & = \frac{1}{N_f} \sum_{i = 1}^{N_f} r_f^2 (\vec{z}_f^i,\theta)  \\ 
J_{BC}(\theta) & = \frac{1}{N_{BC}} \sum_{i = 1}^{N_{BC}} r_{BC}^2 (\vec{z}_{BC}^i,\theta) \\
J_{IC}(\theta) & =  \frac{1}{N_{IC}} \sum_{i = 1}^{N_{IC}} r_{IC}^2 (\vec{x}_{IC}^i,\theta) , 
\end{split}
\end{equation}
and the residuals of the Dirichlet and Neumann boundary conditions are expressed as $r_{BC} (\vec{z},\theta) = \mathcal{B}(\hat{u}(\vec{z},\theta)) - g_{BC} (\vec{z})$, $(\vec{z} \in  \partial \Omega \times (0, T))$ and $\mathcal{B}$ and $g_{BC}$ are the operator and source term associated with the given boundary condition, respectively. The weights $w_{f}$, $w_{BC}$, and $w_{IC}$ penalize the loss terms associated with the governing PDEs, boundary, and initial conditions, respectively. 
In Eq (\ref{eq:loss_components}), $\{ \vec{z}_{BC}^i \}_{i=1}^{N_{BC}}$ are the locations where the boundary conditions are enforced, $\{ \vec{x}_{IC}^i \}_{i=1}^{N_{IC}}$ are the locations where the initial conditions are enforced, and $\{ \vec{z}_{f}^i \}_{i=1}^{N_{f}}$ is the set of \emph{residual} points where the PDE's residuals are minimized. Here, $N_f$ is the number of residual points, $N_{BC}$ is the number of boundary points, and $N_{IC}$ is the number of initial condition points. These points form a set of training points, with the number of training points given by $n_h = N_f + N_{BC} + N_{IC}$. %The loss function (\ref{eq:loss_pinn}) in the PINN method is composed of  the mean square terms arising from the PDE and initial and  boundary conditions. 

 The DNN's  parameters (or weights) $\theta$ are found by minimizing the loss function $J(\theta)$:
 \begin{equation}\label{eq:optimization}
    \theta = \min_{\theta^*} J(\theta^*). 
 \end{equation}
The resulting DNN $\hat{{u}}(\vec{z},\theta)$  satisfies Eq (\ref{eq:ADE}), including the boundary and initial conditions up to the approximation error $|u(\vec{x})-\hat{u}(\vec{x},\theta)|$ and the error in training $\hat{{u}}(\vec{z},\theta)$. 
%
%As the loss $J(\theta)$ is typically minimized by using stochastic gradient descent, a very large number of data and collocation points can be used within each  gradient descent iteration.
It was shown in \cite{Shin2020} for some classes of PDEs  that if the DNN is large enough to approximate the solution (i.e., the approximation error is negligible), the PINN solution of a PDE (i.e., a solution obtained by minimizing the PDE residuals, including the residuals of the boundary and initial conditions) converges to the strong solution of the PDE as the number of training points $n_h$ approaches infinity.

% * Boundary
\emph{Remark}: %To ensure a well-posed problem, initial and boundary conditions are necessary.
For a finite $n_h$, properly selecting the $\omega_f$, $\omega_{BC}$, and $\omega_{IC}$ weights is a critical step to obtain an accurate solution. 
%In conventional discretization-based methods, the difficulites arise from the errors 
The loss function (\ref{eq:loss_pinn}) enforces PDE and initial and boundary conditions as penalty terms rather than hard constraints.  
This approach is commonly used in the PINN  literature~\cite{Raissi2019,Lu2019,Raissi2020,Mao2020,Tartakovsky2020a,he2020physics,Sirignano2018,Weinan2018} because solving unconstrained optimization problems such as Eq (\ref{eq:optimization}) is in general easier than solving a constrained optimization problem. Also, the solution accuracy can be improved by increasing the number of training points, $N_{BC}$, $N_{IC}$, and $N_f$.
%  $N_{BC}$ and $N_{IC}$. 
In this work, we investigate the effect of the relative values of the weights $\omega_f$, $\omega_{BC}$, and $\omega_{IC}$ on the accuracy of the solution and show that accurate solutions of ADEs can usually be achieved when $\omega_{BC}$ and $\omega_{IC}$ are larger than $10\omega_f$.

Another approach for imposing boundary conditions in solving boundary value problems includes the change of variables where a function that exactly satisfies the boundary conditions is introduced to construct a composite DNN~\cite{Lagaris1997,Lagaris2000a,Yadav2016}.
Originally, this approach was limited to simple domain geometries and linear boundary conditions, but it has recently been  extended to complex boundary geometries by using the level set method and additional neural networks that are separately trained to satisfy the boundary conditions~\cite{Berg2018,Zhu2020}. Extending this approach to non-homogeneous, non-linear boundary conditions still remains a challenge.

\subsection{Training algorithm}
%
% * Training algorithm
The optimization problem (\ref{eq:optimization}) is non-convex.  
%The properties of DNNs for function approximation (including the solutions of PDEs) are well understood  ~\cite{grohs2018proof,Shin2020,Hornik1989,cybenko1989approximation} -- in the limit of the infinite width of the hidden layers, a DNN can exactly approximate any function. 
%Nonconvex optimization remains an active research area, and the optimization errors in the PINN method are poorly understood.  
%
%For gradient-based training of deep neural netowrks, optimization often involves many engineering tricks and tedious fine-tuning of hyper-parameters~\cite{Bengio2012}. 
Common methods for minimizing loss functions in DNN training with and without physics constraints include  the stochastic-gradient descent  Adam method~\cite{Kingma2014} and the limited memory BFGS with box constraints (L-BFGS-B)~\cite{Byrd1995}. We find that the errors in the PINN method strongly depend on which of these two algorithms is used. Therefore, in this work, we use a two-step optimization algorithm that was found to perform well in the application of the PINN method for parameter estimation ~\cite{Lu2019,he2020physics}. 
In this two-step algorithm, the Adam method is used first for a prescribed number of iterations, and then the L-BFGS-B method is used until the solution of the optimization problem converges with the prescribed tolerance.  At the beginning of the Adams step, the DNN weights are randomly initialized using the Xavier scheme~\cite{Glorot2010a}.

%%%%%%%%%%%%%%%%%%%%%%%%%%%%
%%% Numerical model & PINN %%%
%%%%%%%%%%%%%%%%%%%%%%%%%%%%

\section{Forward flow and advection-dispersion equations}
\label{sec:result}
%
% * review the examples
Here, we present four numerical experiments to demonstrate the effectiveness of the PINN approach for solving forward ADEs.

First, we  compare the PINN solutions of the 1D (Section \ref{sec:1D_AD}) and 2D (Section \ref{sec:2Dunsteady}) time-dependent ADEs with the analytical solutions that are commonly used to benchmark numerical grid-based methods \cite{Hughes2005,Huang2008,Borker2017,Mojtabi2015}. In Section \ref{sec:grid_effect}, we investigate the effect of grid orientation on the ADE solution with $Pe \gg 1$, where the crosswind diffusion could lead to numerical instabilities in discretization-based methods. 
%For problems with large  $Pe$, traditional numerical methods might require high-order time-integration schemes and fine mesh. The objective of these tests is to demonstrate the capability of the PINN approach to solve problems for $Pe. 

In Section \ref{sec:2DSTOMP}, we consider a two-dimensional system of the steady-state Darcy flow equation with heterogeneous conductivity field and a time-dependent ADE with the anisotropic dispersion coefficient. This problem presents a  challenge for numerical approximations of off-diagonal entries in the dispersion tensor, often resulting in nonphysical negative solutions \cite{herrera2006positive,lipnikov2007monotone,herrera2010multidimensional}. This example aims to show that PINN can provide comparable solutions to state-of-art numerical solvers such as the Subsurface Transport Over Multiple Phase (STOMP) finite volume code~\cite{White1995} for ADEs with anisotropic dispersion coefficients and non-uniform velocity fields.

The quantitative comparison of PINN solutions with the reference (analytical or grid-based numerical) solutions is given in terms of the relative $L_2$ error
\begin{equation}\label{eq:error}
\epsilon = \frac{||\vec{u} - \hat{\vec{u}}||}{||\vec{u}||}
\end{equation}
that we compute on a uniform grid over the space-time domain,
where the vectors $\vec{u}$ and $\hat{\vec{u}}$ denote the reference solution and the PINN solution (evaluated at the grid points), respectively. 

We define the DNN size as $n_l \times m_l$, where $n_l$ is the number of hidden layers and $m_l$ is the number of neurons in each hidden layer.
For the time-dependent two-dimensional problems considered in the study, we use the $n_t \times n_1 \times n_2$ notation to define $N_f$, $N_{BC}$, and $N_{BC}$ in (\ref{eq:loss_components}). In this notation, $n_t$ denotes the number of time steps along the temporal coordinate (including the initial time), $N_{IC} = n_1 \times n_2$, and $N_{BC} = n_t \times 2 (n_1+n_2)$. Unless stated otherwise, the number of residual points is $N_f = n_t \times n_1 \times n_2$.
The residual data points are randomly distributed over the space-time domain.

%%% ---------------------------------
\subsection{One-dimensional time-dependent ADE}\label{sec:1D_AD}

\begin{figure} [ht!]
	\captionsetup[subfloat]{farskip=0.0pt,captionskip=0pt}
	\centering
	\subfloat[$t=0.8$] {\includegraphics[angle=0,width=1.6in]{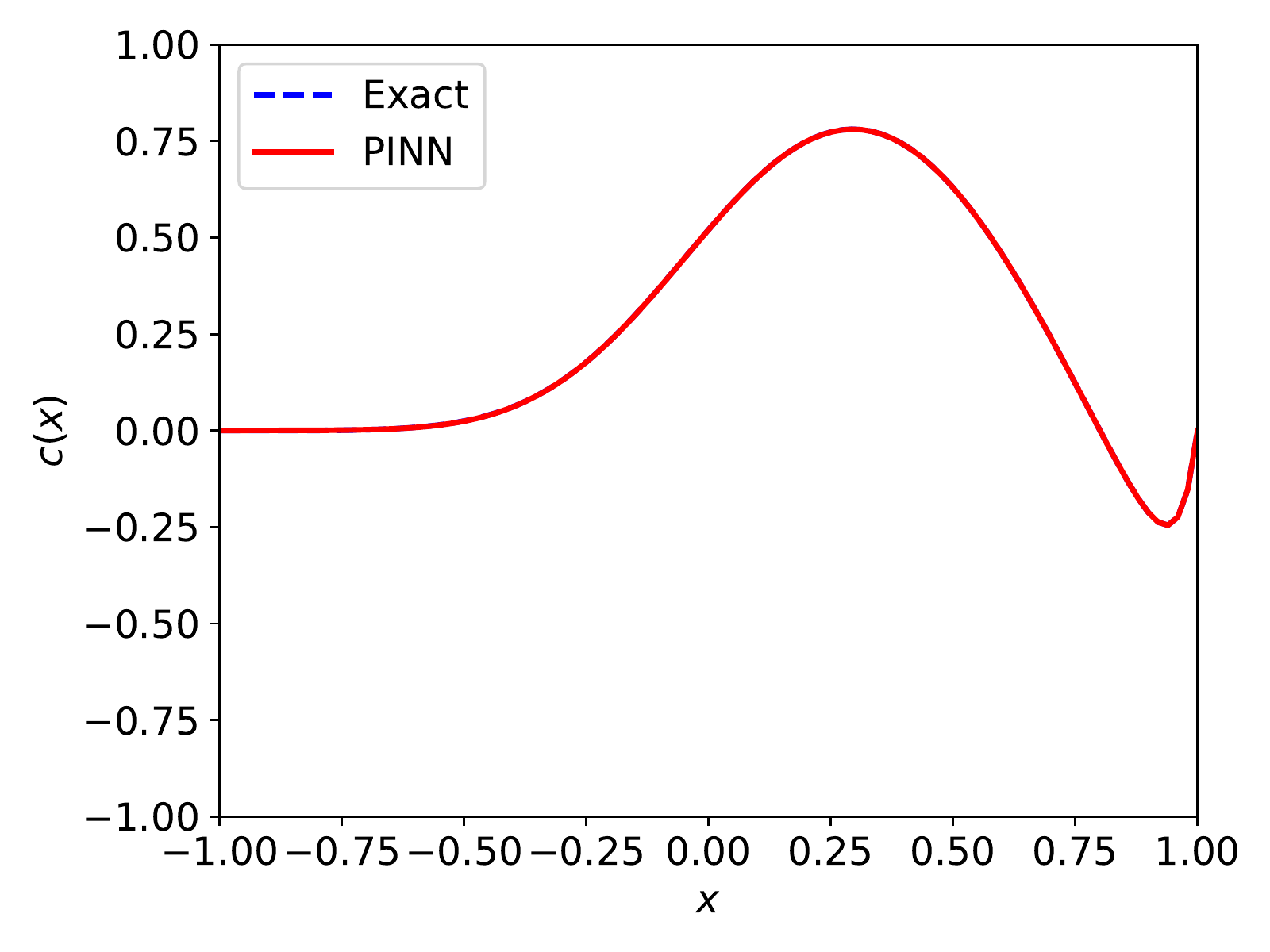}}
	\subfloat[$t=1.0$] {\includegraphics[angle=0,width=1.6in]{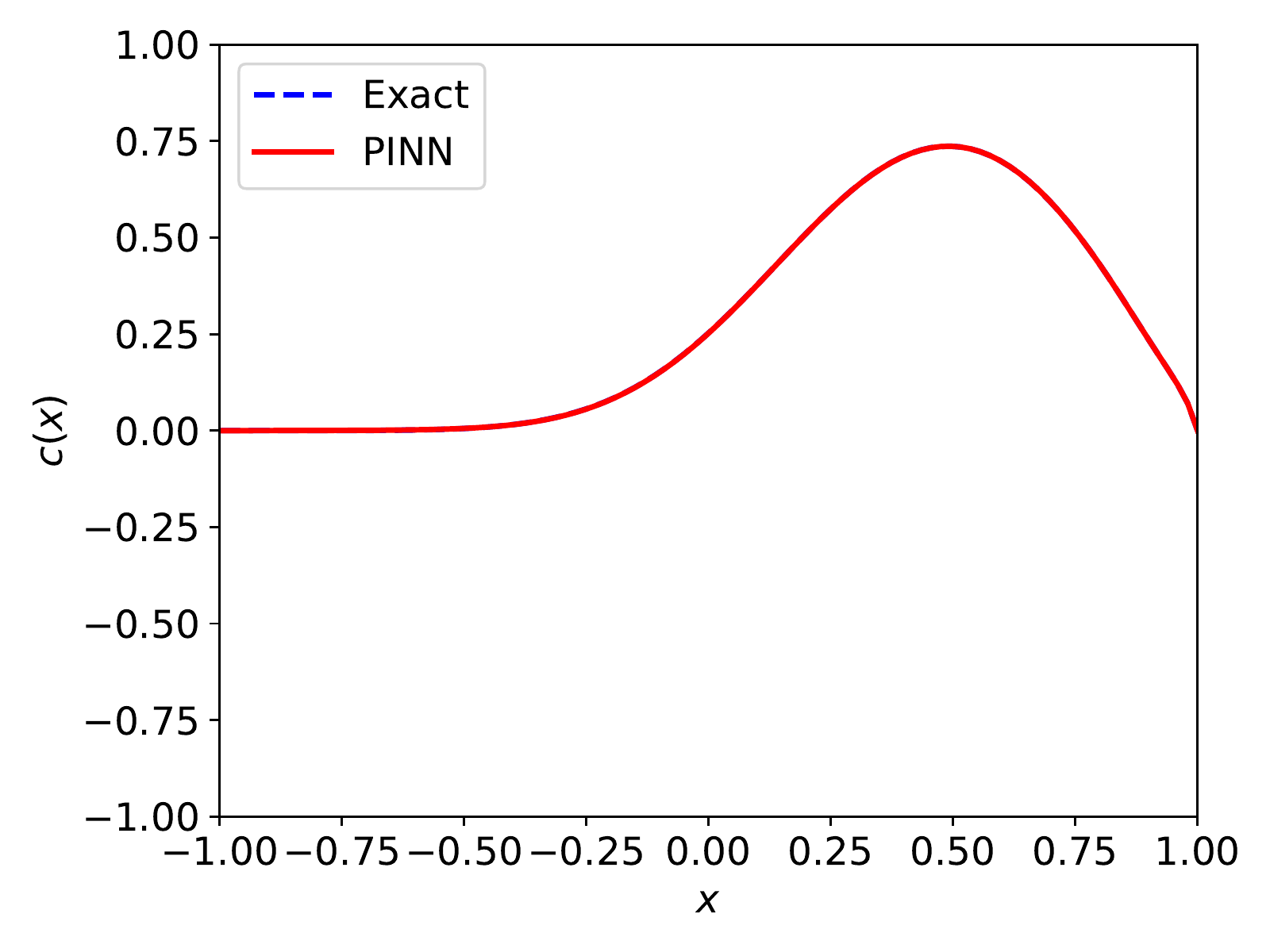}}
	\subfloat[$t=1.6$] {\includegraphics[angle=0,width=1.6in]{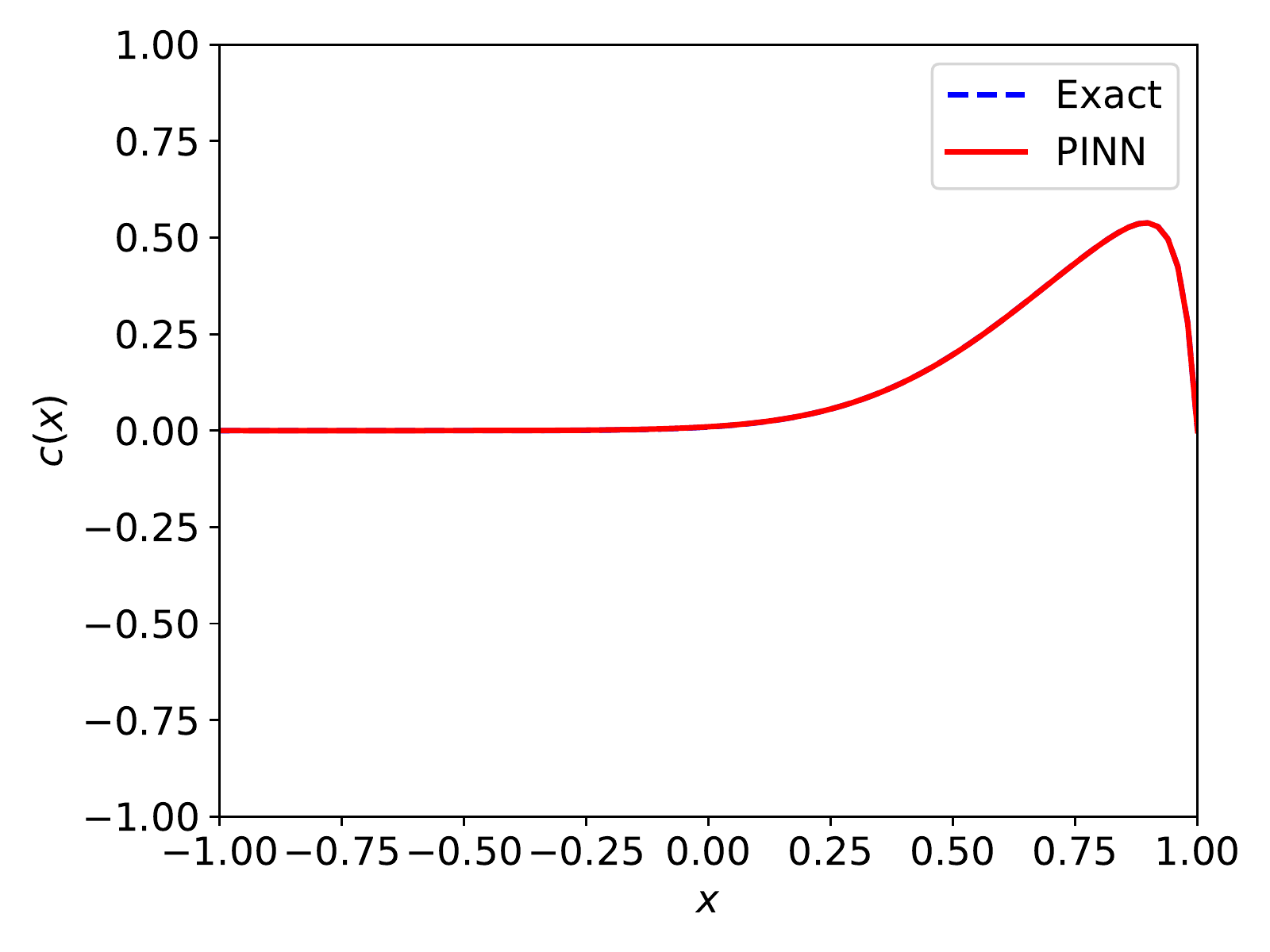}}
	\caption{Comparison of the PINN and analytical solutions of the time-dependent ADE (\ref{eq:AD1d_pde}) as functions of $x$ at times $t=0.8, 1.0$, and $1.6$. $Pe = 62.8$, and the DNN size is $4 \times 40$.}
	\label{fig.AD1d_sin_soln}
\end{figure}

\begin{figure} [ht!]
	\captionsetup[subfloat]{farskip=0.0pt,captionskip=0pt}
	\centering
	\subfloat[$t=0.8$] {\includegraphics[angle=0,width=1.6in]{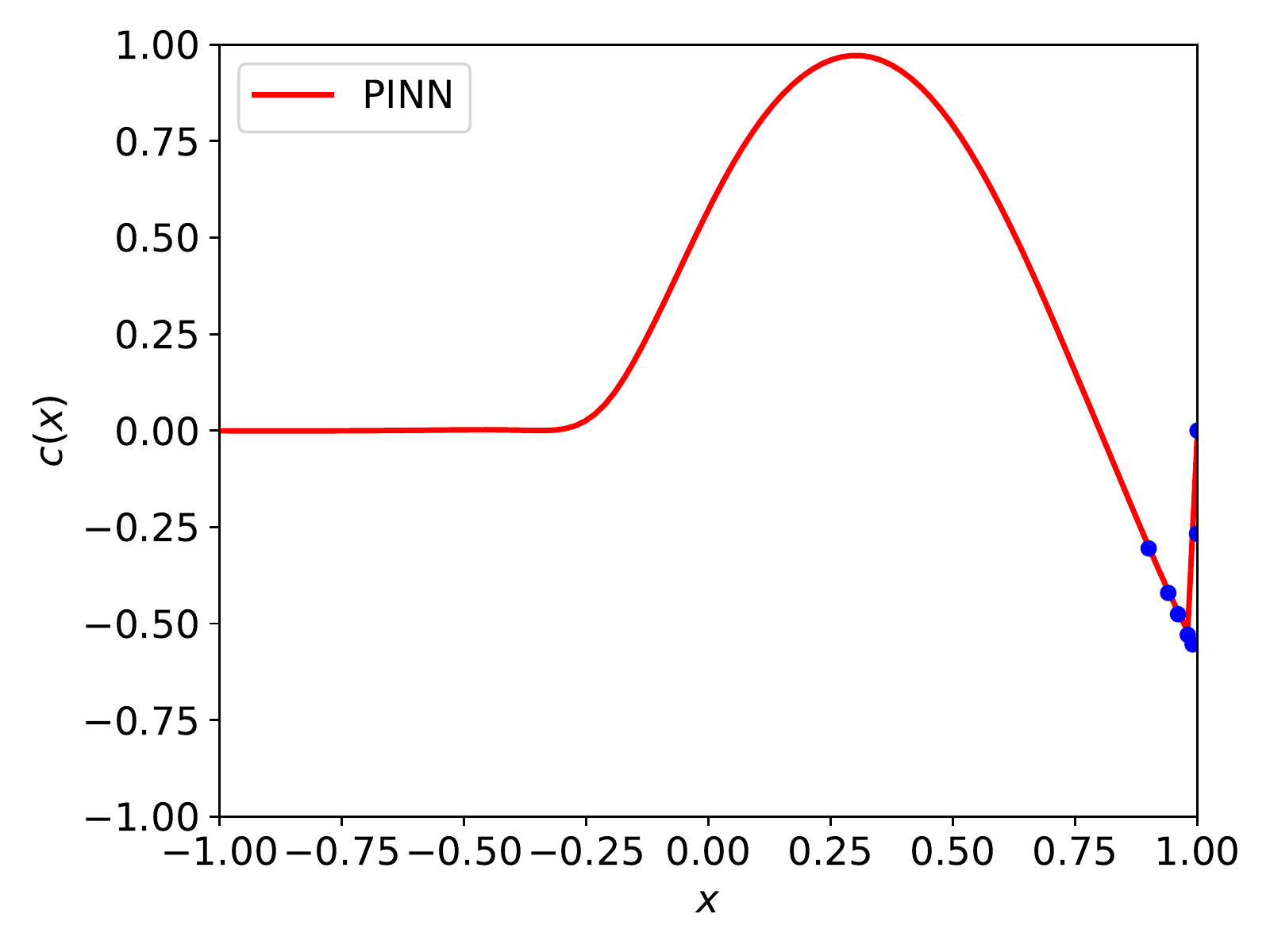}}
	\subfloat[$t=1.0$] {\includegraphics[angle=0,width=1.6in]{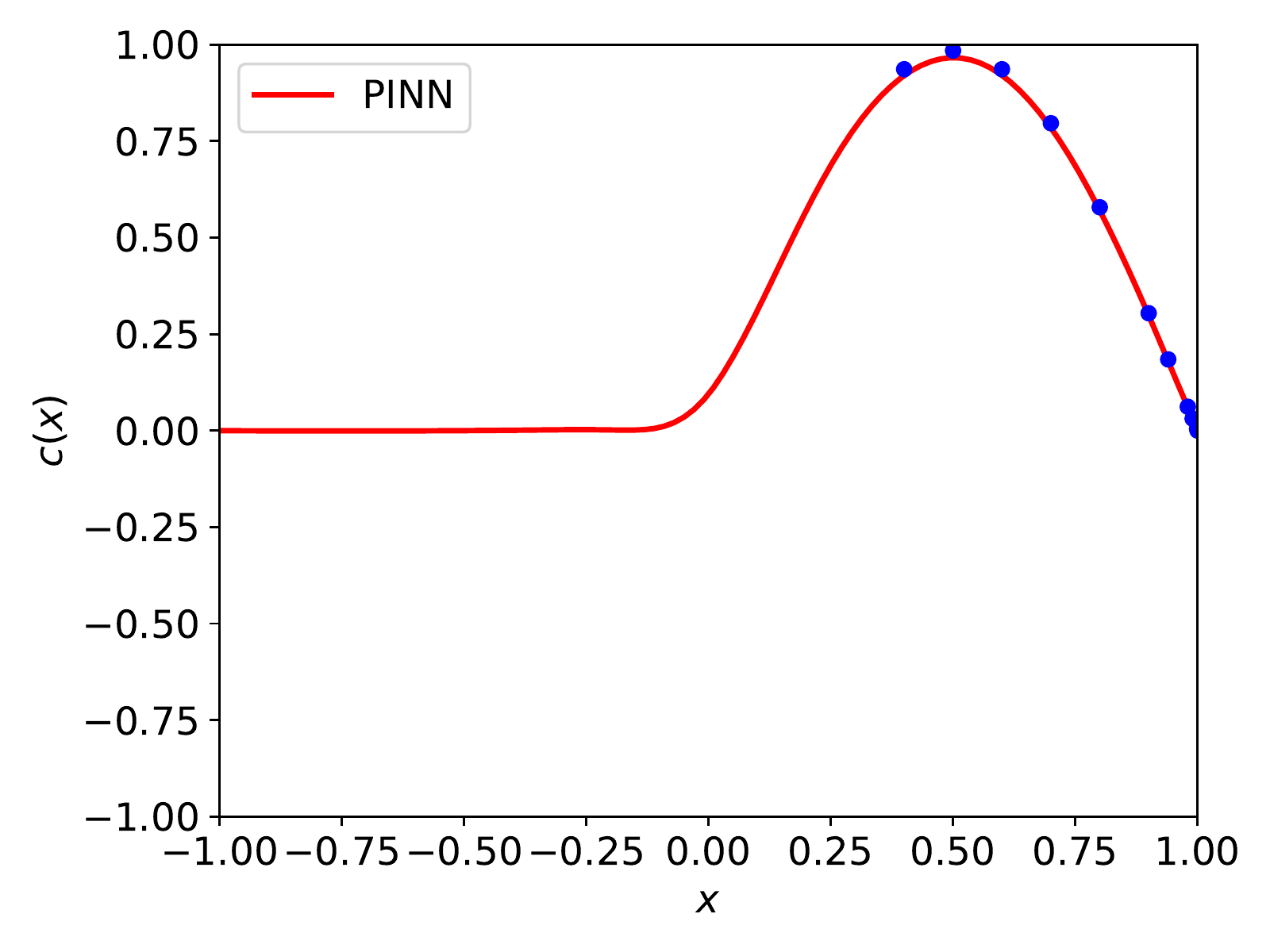}}
	\subfloat[$t=1.6$] {\includegraphics[angle=0,width=1.6in]{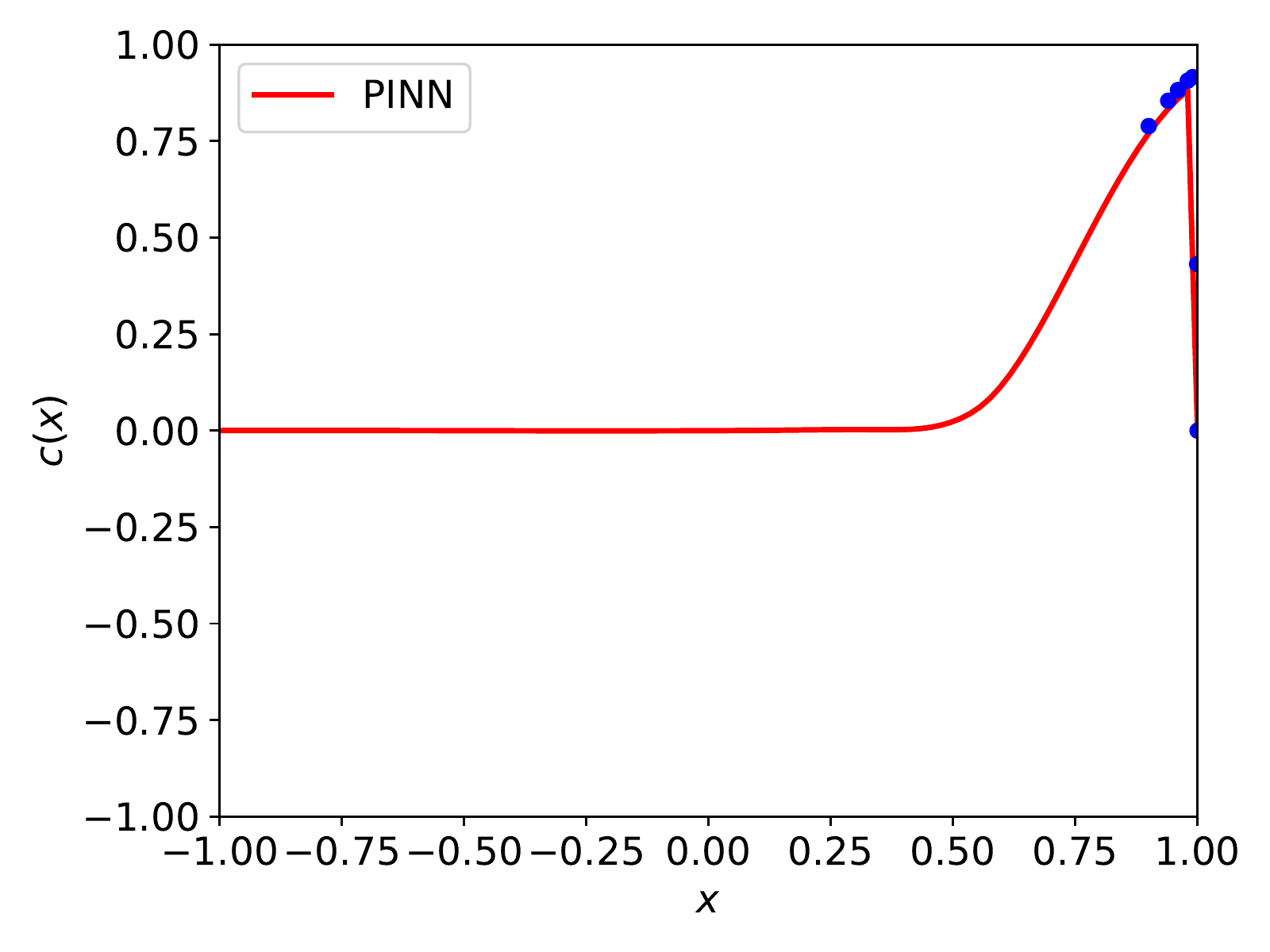}}
	\\
	\subfloat[$t=0.8$] {\includegraphics[angle=0,width=1.6in]{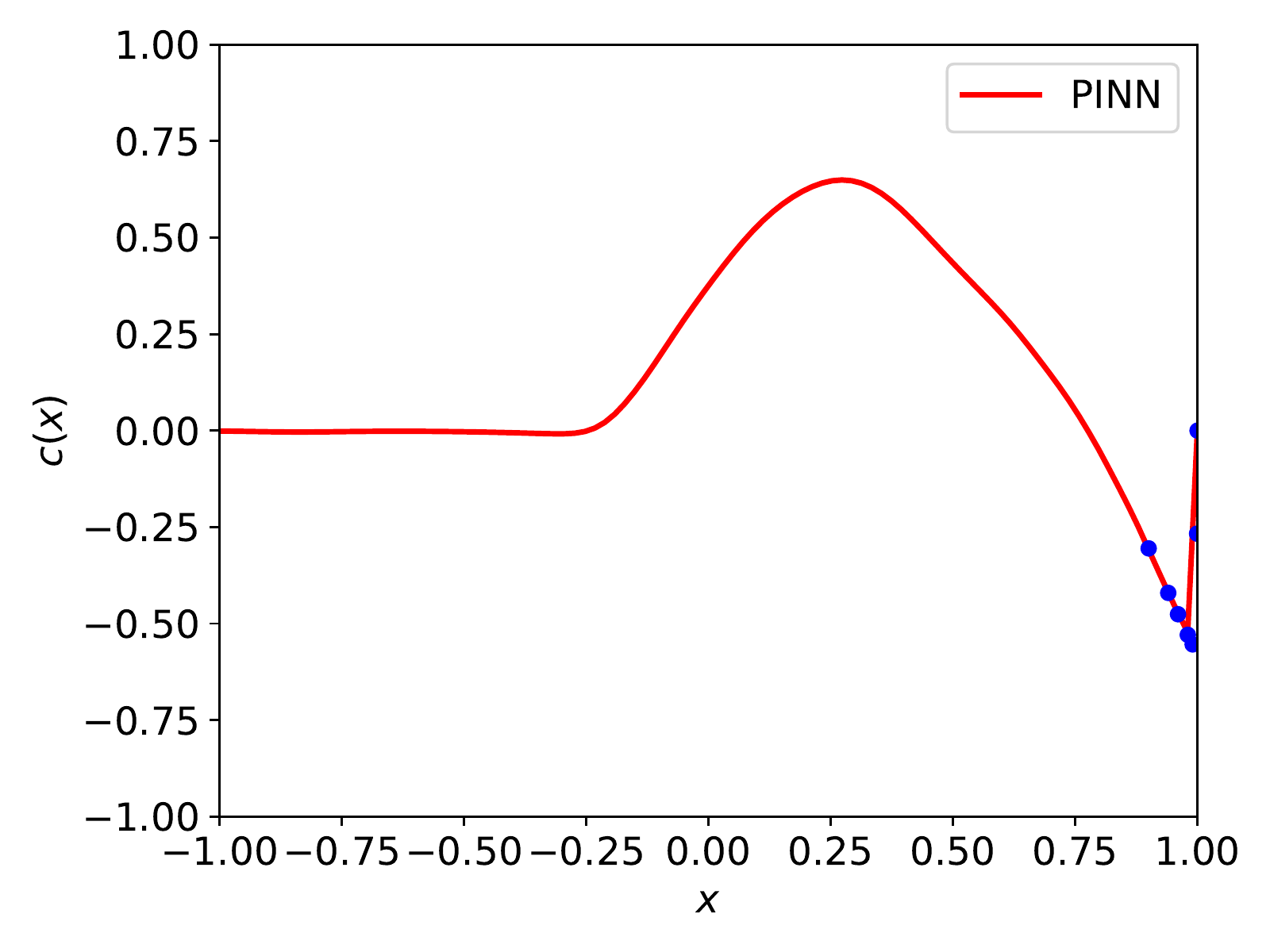}}
	\subfloat[$t=1.0$] {\includegraphics[angle=0,width=1.6in]{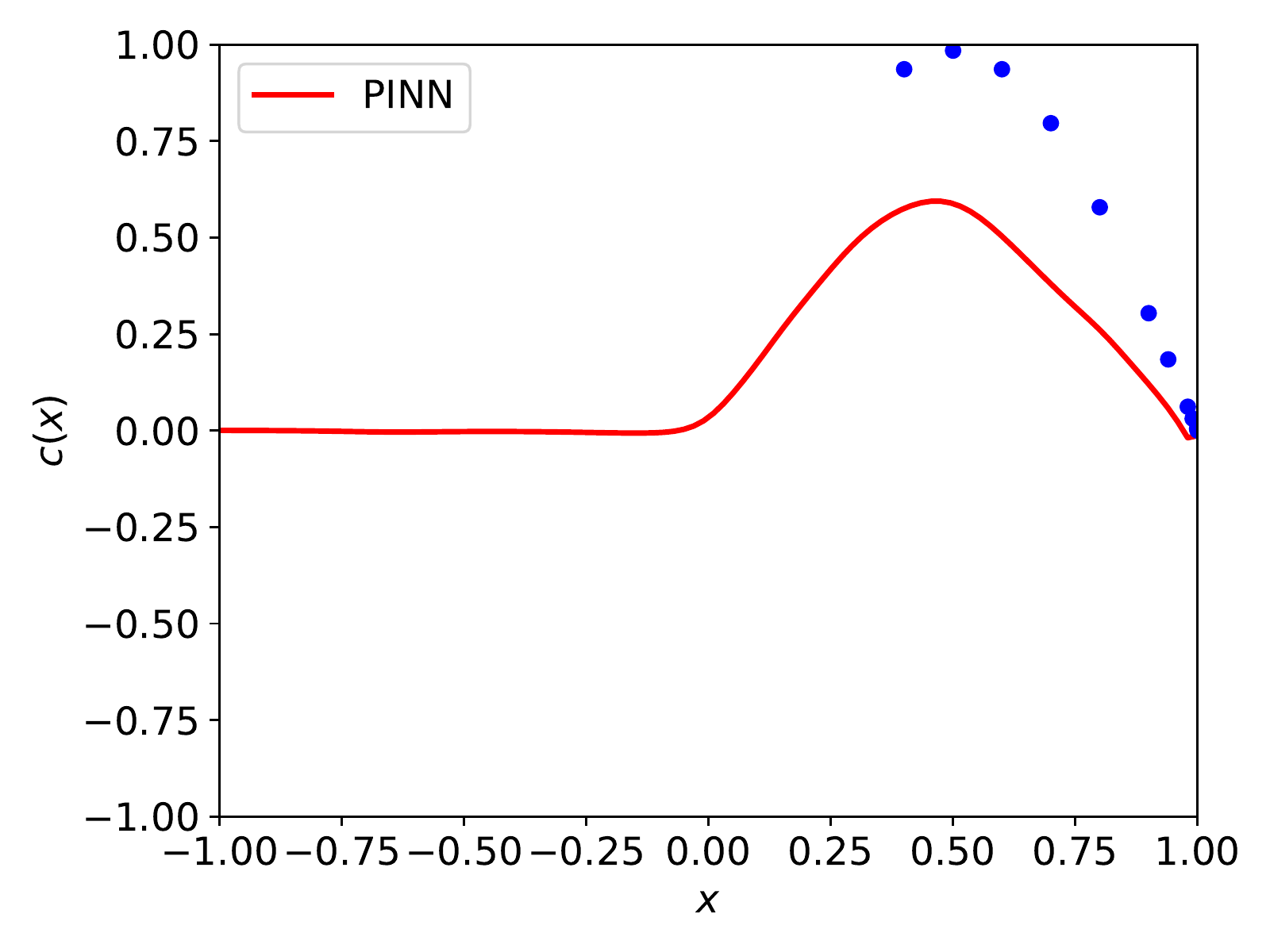}}
	\subfloat[$t=1.6$] {\includegraphics[angle=0,width=1.6in]{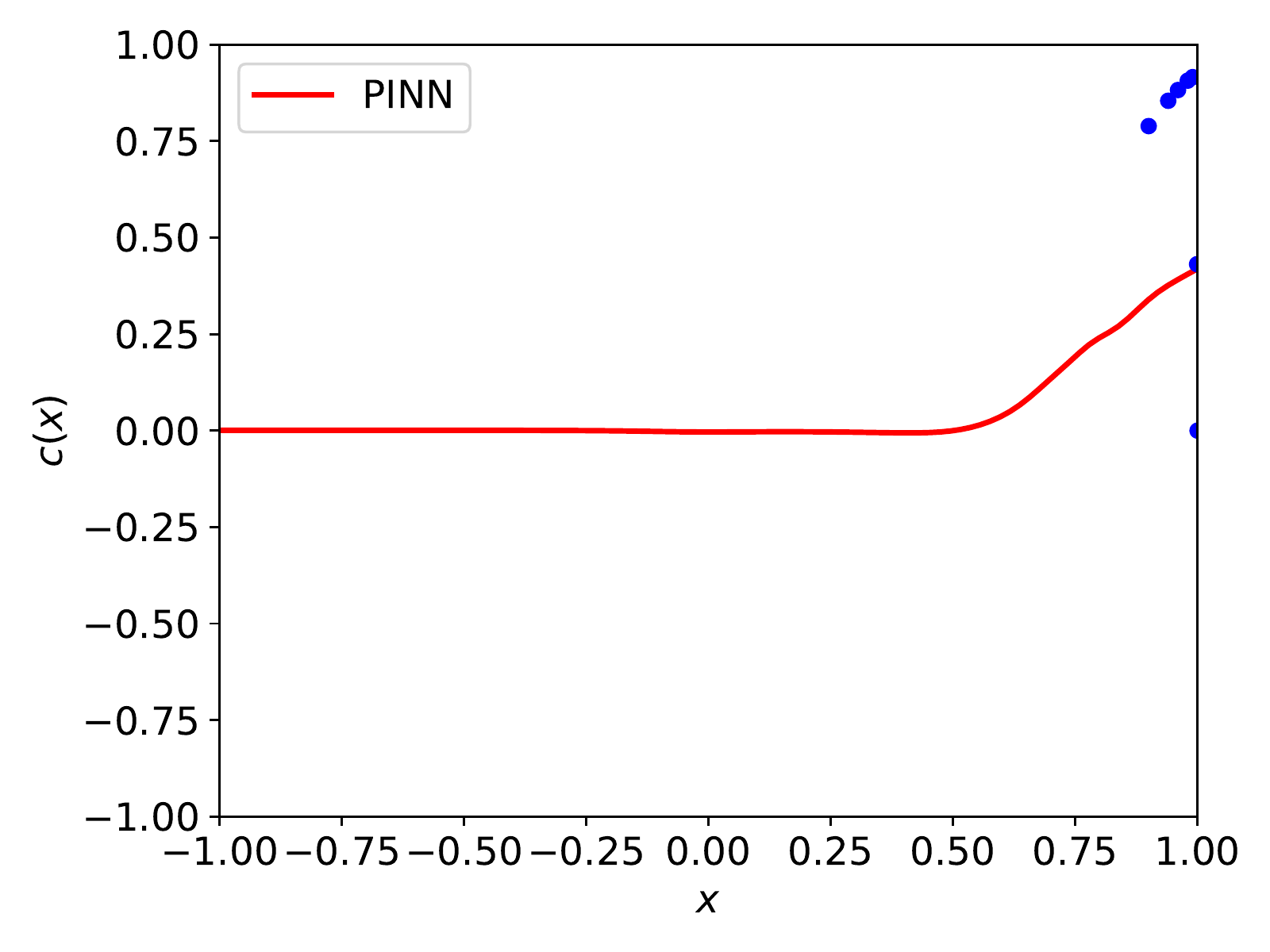}}
	\caption{Comparison of the PINN and analytical solutions of the time-dependent ADE (\ref{eq:AD1d_pde}) as functions of $x$ at times  $t=0.8$, $1.0$, and $1.6$. $Pe = 628$. PINN solutions in (a)--(c) are obtained with the DNN size $4 \times 40$ and in (d)--(f) with the DNN size $3 \times 40$. The approximate analytical solution (blue dots) is obtained in \cite{Mojtabi2015} and is valid for $x > at -1$.}
% 	\textcolor{red}{finite element (or analytical?)}\textcolor{blue}{finite element. Due to the high Peclet number, the analytical is not accurate for this case. Thus, I collected the solution points provided in other literature.} solution given by~\cite{Mojtabi2015}. \textcolor{red}{Why the blue dots are only shown for large $x$? }\textcolor{blue}{QH: they are some point values provided in Paper~\cite{Mojtabi2015}, and I only selected some of them for comparison. It is consistent to the Figure 2 in \cite{Khodayi-Mehr2019} }
	\label{fig.AD1d_sin_soln_K001}
\end{figure}

Consider the following one-dimensional ADE ~\cite{Mojtabi2015,Khodayi-Mehr2019} 
%defined for $T=[0,2]$ and $\Omega=[-1,1]$:
\begin{equation}\label{eq:AD1d_pde}
\frac{\partial u}{\partial t} + a \frac{\partial u}{\partial x} = \kappa \frac{\partial^2 u}{\partial x^2}, \quad -1<x<1, \quad t>0 
\end{equation}
with the initial and boundary conditions:
\begin{equation}\label{eq:AD1d_bc}
\left\{
\begin{array}{ll}
\begin{split}
& u(x,t=0) = -\sin(\pi x), \quad -1<x<1 \\
& u(x=0,t)  = 0,\quad u(x=1,t) = 1, \quad  t > 0,
\end{split}
\end{array} \right.
\end{equation}
%$\kappa = 0.01/\pi, 0.08/\pi, 0.1/\pi, 0.2/pi$
%shown in \textcolor{blue}{Table} XXX 
where the velocity is set to $a = 1$. For this problem, we study the accuracy of the PINN method for different $Pe = l\times a/\kappa$, where $l=2$ is the domain size, against the analytical solution~\cite{Mojtabi2015}:
\begin{equation}\label{eq:AD1d_ExactSoln}
\begin{split}
& u (x,t)  = 16 \pi^2 \kappa^3 a e^{\frac{c}{2 \kappa}(x-\frac{c}{2}t)} \\
& \times \left[  \sinh {\left(  \frac{a}{2 \kappa}  \right)}
\sum_{p=0}^{N=\infty}   \frac{(-1)^p 2 p \sin (p \pi x) e^{-\kappa p^2 \pi^2 t}}{a^4 + 8 (a \pi \kappa)^2 (p^2 +1) + 16 (\pi \kappa)^4 (p^2-1)^2 }\right. \\
& \left. +  \cosh{ \left(  \frac{a}{2 \kappa}  \right)}  \sum_{p=0}^{N=\infty} \frac{(-1)^p (2 p +1) \cos \left( \frac{2p+1}{2} \pi x\right) e^{-\kappa \frac{(2p+1)^2}{4} \pi^2 t}}{a^4 + (a \pi \kappa)^2 (8p^2 +8p +10) + (\pi \kappa)^4 (4p^2 +4p -3)^2 }	 \right].
\end{split}
\end{equation}
Here, we use $N=800$ to evaluate the analytical solution. 

The PINN solution is obtained on the time domain $T=[0,2]$.  
We enforce the initial condition at $N_{IC} = 100$ points, boundary condition at $N_{BC} = 200$ points, and minimize the PDE residual at $N_f = 200 \times 100$ collocation points.
For the two-step training scheme, we set the number of Adam iterations to $20,000$ with a learning rate of $0.001$. The mini-batch size of $500$ is adopted such that $500$ residual points are randomly selected from $\{(\vec{x}_{f}^i,t_{f}^i)\}_{i=1}^{N_{f}}$ within each Adam iteration. 

%optimizer when the total loss reaches a prescribed value of = 5 × 10−4 and apply the L-BFGS-B optimizer to achieve the final convergence of the loss function. The stochastic gradi- ent descent method in the Adam algorithm causes oscillations in losses that allows a better DNN generalization. The quasi-Newton L-BFGS-B method enables a higher rate of convergence to the minimum identified by the Adam algorithm. The small final value of indicates that the,

The snapshots of the PINN and analytical solutions as functions of $x$ at $t=0.8$, 1.0, and 1.6 for $Pe = 62.8$ ($\kappa = 0.1/\pi$) are given in Figure \ref{fig.AD1d_sin_soln}. The PINN solution is obtained with a $4 \times 40$ DNN and has a near-perfect agreement with the analytical solution, with the relative error $\epsilon = 8.92 \times 10^{-4}$. Table \ref{table:AD1Dt_sin_comp} gives $\epsilon$ as a function of the DNN size for this case. The error can be reduced by increasing both the number of hidden layers and the width of the layers. For all considered cases, however, the error stays below 1\%. 

Figure \ref{fig.AD1d_sin_soln_K001} shows the PINN and analytical solutions for $Pe=628$ ($\kappa = 0.01/\pi$)  as functions of $x$ at three different times. In Figures \ref{fig.AD1d_sin_soln_K001} (a)--(c), the PINN solution is obtained with a $4\times 40$ DNN, while  the solution in  Figures \ref{fig.AD1d_sin_soln_K001} (e)--(d) is obtained with a smaller $3\times40$ DNN. 
Note that the general analytical solution (\ref{eq:AD1d_ExactSoln}) develops oscillations for large $Pe$  (including the $Pe = 628$ considered here) and that an analytical solution obtained by a perturbed wave equation in \cite{Mojtabi2015} for the region $x > at -1 $ is used in Figure \ref{fig.AD1d_sin_soln_K001} as a reference. 
We can see that for this $Pe$, the DNN size plays a more significant role than for $Pe=62.8$. The solution with  $4\times 40$ DNN provides a perfect fit with the analytical solution, while the solution with the smaller DNN has a maximum point error close to 100\%. The reason for this is that the solution develops very large gradients near the $x=1$ boundary, and a sufficiently large DNN is needed to accurately approximate it. 
We note that in numerical discretization-based solutions of ADEs, an increase in $Pe$ often requires a finer mesh to maintain the same accuracy \cite{Yadav2016,Borker2017}.

\begin{figure}[!ht]
\captionsetup[subfloat]{farskip=0.0pt,captionskip=0pt}
\centering
\subfloat[] {\includegraphics[angle=0,width=2.4in]{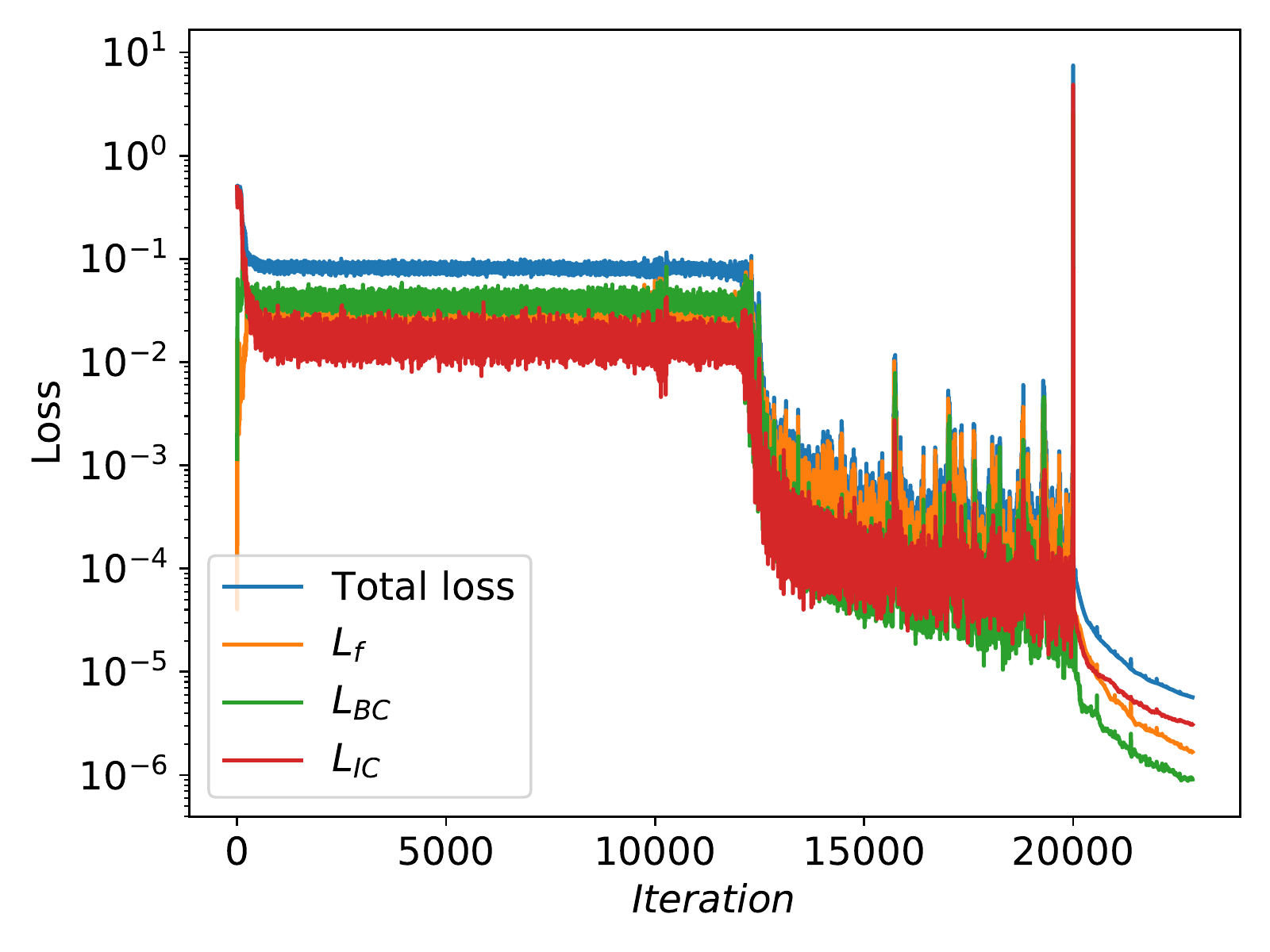}}
\subfloat[] {\includegraphics[angle=0,width=2.4in]{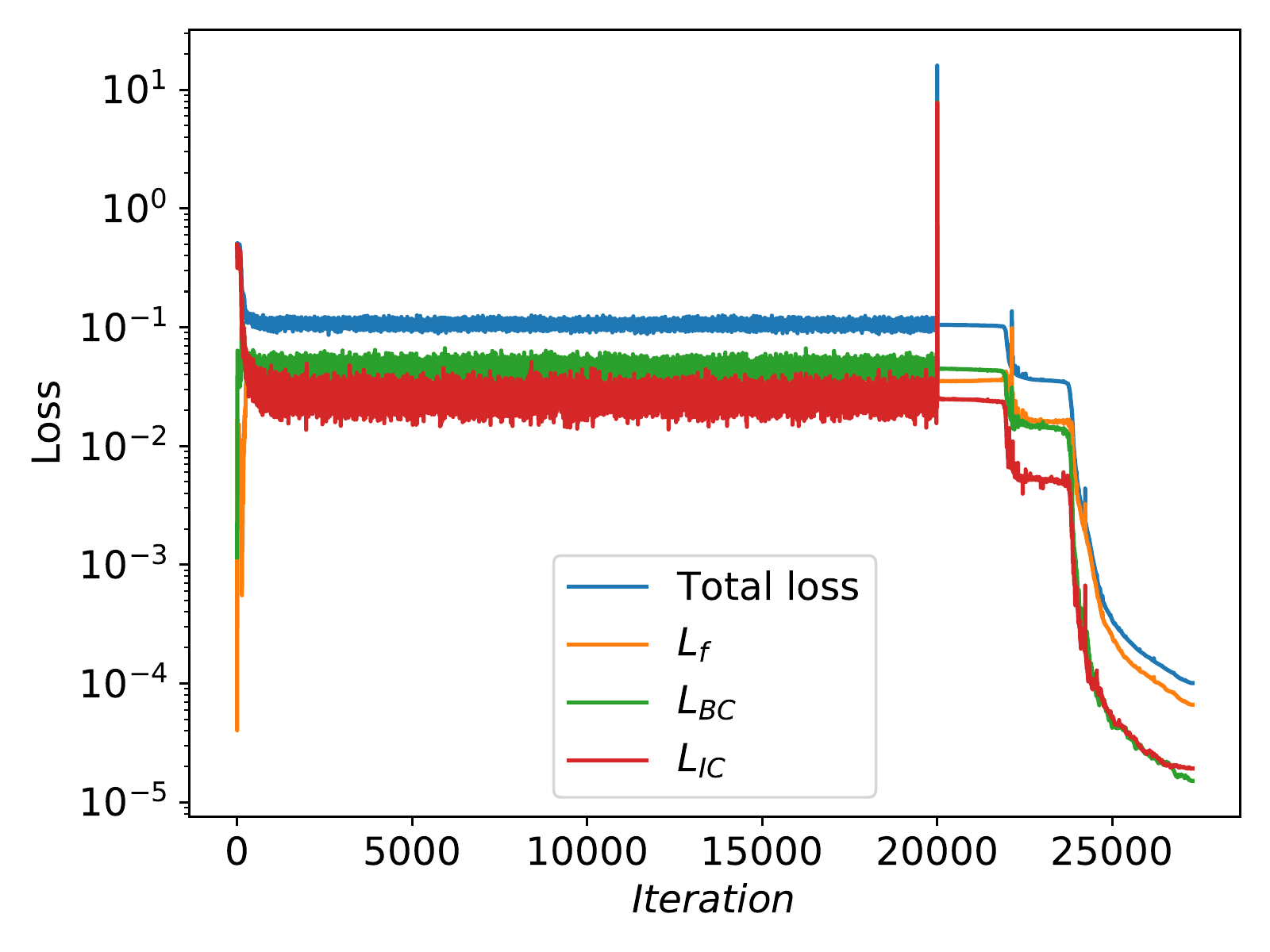}}
\caption{Loss functions for the ADE (\ref{eq:AD1d_pde}) with: (a) $\kappa=0.1 /\pi$ ($Pe=62.8$) and (b) $\kappa=0.01 /\pi$ ($Pe=628$). The DNN size $4 \times 40$ is used for the PINN approach.}
\label{fig:loss_AD1d}
\end{figure}

% * Loss
Figure \ref{fig:loss_AD1d} shows the total loss and the individual loss terms versus the number of iterations by using the two-step training scheme for $Pe=62.8$ and 628. In this optimization scheme, the Adam optimizer terminates at the prescribed number of iterations ($20,000$) and switches to the L-BFGS-B optimizer to achieve the final convergence. The mini-batch stochastic gradient descent-based Adam algorithm has a good overall performance but produces relatively large oscillations at the late stage of the training process. Using the L-BFGS-B optimizer at the second stage of training reduces oscillations and increases the convergence rate.  On the other hand, the use of L-BFGS-B alone might result in the training being trapped in a local minimum corresponding to a relatively large value of the loss function. 
The comparison of Figures \ref{fig:loss_AD1d}a ($Pe=68.2$) and \ref{fig:loss_AD1d}b ($Pe = 628$) demonstrates that the final loss value increases with increasing $Pe$, which is mainly due to the larger final value of the residual loss ($L_f$). Also, the number of iterations required to achieve the same tolerance increases with increasing $Pe$. For these two examples, it takes nearly $5,000$ more iterations during the quasi-Newton L-BFGS-B for $Pe=628$ than for $Pe=62.8$. 
% We note that a smaller learning rate in Adam could be used to reduce the training error.
%However, by using the two-step training scheme, good convergence are still obtained for both cases, as evidenced by their solutions in Figure \ref{fig.AD1d_sin_soln} and \ref{fig.AD1d_sin_soln_K001}a, respectively.

\begin{table}[htb]
	\centering
	\small
	\caption{Relative $L_2$ error $\epsilon$ for the ADE (\ref{eq:AD1d_pde}) with $\kappa=0.1 /\pi$ ($Pe = 62.8$). The error is calculated on a uniform $100 \times 100$ grid over the space-time domain $\Omega \times T$.}
	%	\resizebox{\columnwidth}{!}{%}
	\begin{tabular}{cccc}
		\toprule
		DNN size & $\varepsilon$ \\
		\hline
		\multicolumn{1}{c}{$2 \times 30$}  & $9.186 \times 10^{-3}$   \\
		\multicolumn{1}{c}{$2 \times 50$}  & $4.152 \times 10^{-3}$   \\	
		\multicolumn{1}{c}{$3 \times 30$}  & $1.676 \times 10^{-3}$   \\	
		\multicolumn{1}{c}{$3 \times 50$}  & $1.218 \times 10^{-3}$   \\	
		\multicolumn{1}{c}{$4 \times 30$}  & $5.854 \times 10^{-4}$   \\	
		\multicolumn{1}{c}{$4 \times 50$}  & $6.381 \times 10^{-4}$   \\	
		\bottomrule
	\end{tabular}%
	\label{table:AD1Dt_sin_comp}
\end{table}

% We note that in~\cite{Khodayi-Mehr2019}, the same equation was solved with 
%  VarNet, a similar to PINN approach, but with significantly larger errors that we  report here. We attribute a better performance of the proposed PINN approach to the better choice of weights in the loss function and the two-step training algorithm. 

%%% ---------------------------------
% 2Dt
%%% ---------------------------------
%
\subsection{Two-dimensional time-dependent ADE:  instantaneous Gaussian source}\label{sec:2Dunsteady}
% 2020_Dwivedi; 2017_Borker, Farhat, et al.
Here, we study the performance of the PINN method for the ADE:
\begin{equation}\label{eq:2DADE}
\left\{
\begin{array}{ll}
\begin{split}
& u_t + \nabla \cdot (-\kappa \nabla u + \vec{a} u) = 0, \quad \vec{x}\in \Omega =(0,1) \times(0,1) \quad t\in (0, T)\\
& \vec{a}= [\cos(\phi) \; \sin(\phi)]^T \quad \phi = 22.5^{\circ} \\
& u  = \frac{1}{4t+1}\exp{\left(-\frac{||\vec{x}-\vec{a}t||^2}{\kappa (4t +1)} \right)}, \quad \vec{x} \in  \partial \Omega\quad t\in (0, T) \\
& u (\vec{x},t=0) = \exp{\left(-\frac{||\vec{x}||^2}{\kappa } \right)}, \quad  \text{in} \quad  \Omega.
\end{split}
\end{array} \right.
\end{equation}
% \textcolor{red}{What is $q$? $q=0$?}
% that was proposed in \cite{Borker2017} for benchmarking numerical methods. 
This equation allows the analytical solution 
\begin{equation}\label{eq:AD2Dt_pts_exact}
u = \frac{1}{4t+1}\exp{\left(-\frac{||\vec{x}-\vec{a}t||^2}{\kappa (4t +1)} \right)}.
\end{equation}

We obtain PINN solutions for $\kappa =  0.02$  and $ 0.005$, corresponding to $Pe=50$ and $200$, on the time domain $[0, 0.6]$ with the residual points spanning the time interval [0,0.5]. The time-space grid is set as $100 \times 40 \times 40$. %and the number of residual points is set to $N_f = 10000$. The residual points are uniformly spaced in the time-space domain including with time spanning the interval [0,0.5].
Here, we adopt the same hyper-parameters as in Section \ref{sec:1D_AD} for the two-step minimization of the loss function.

\begin{figure}[!ht]
	\captionsetup[subfloat]{farskip=0.0pt,captionskip=0pt}
	\centering
	\includegraphics[angle=0,width=5in]{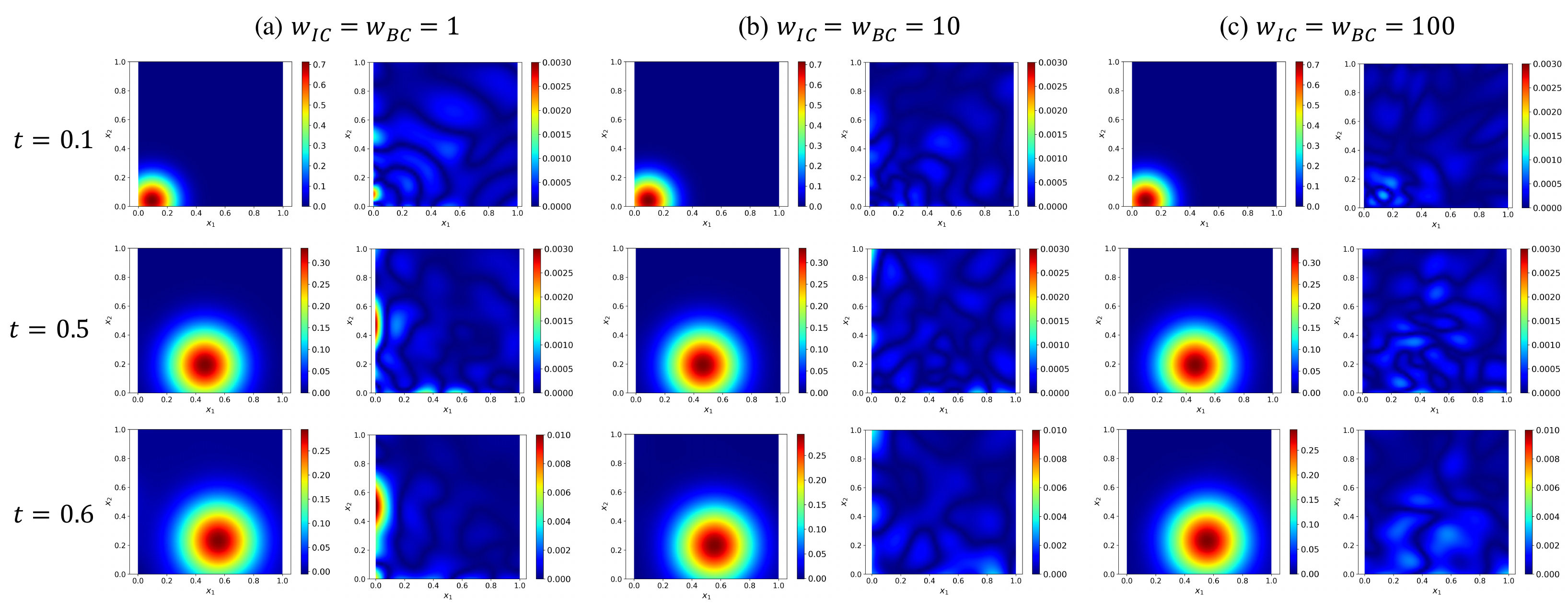}
	\caption{Snapshots of the PINN solution of the time-dependent ADE (\ref{eq:2DADE}) with $Pe = 50$ (left collumns) and the corresponding point errors with respect to the analytical solution (\ref{eq:AD2Dt_pts_exact}) (right columns) obtained with: (a) $w=w_{IC} = w_{BC} = 1$, (b) $w=w_{IC} = w_{BC} = 10$, and (c) $w=w_{IC} = w_{BC} = 100$. For $t = 0.5$, the maximum point errors for $w=1$, $10$, and $100$ are $2.96 \times 10^{-3}$, $1.33 \times 10^{-3}$, and $0.97 \times 10^{-3}$, respectively.}
	\label{fig:AD_2Dt_pts_Pe50}
\end{figure}

\begin{figure}[!ht]
	\captionsetup[subfloat]{farskip=0.0pt,captionskip=0pt}
	\centering
	\includegraphics[angle=0,width=5in]{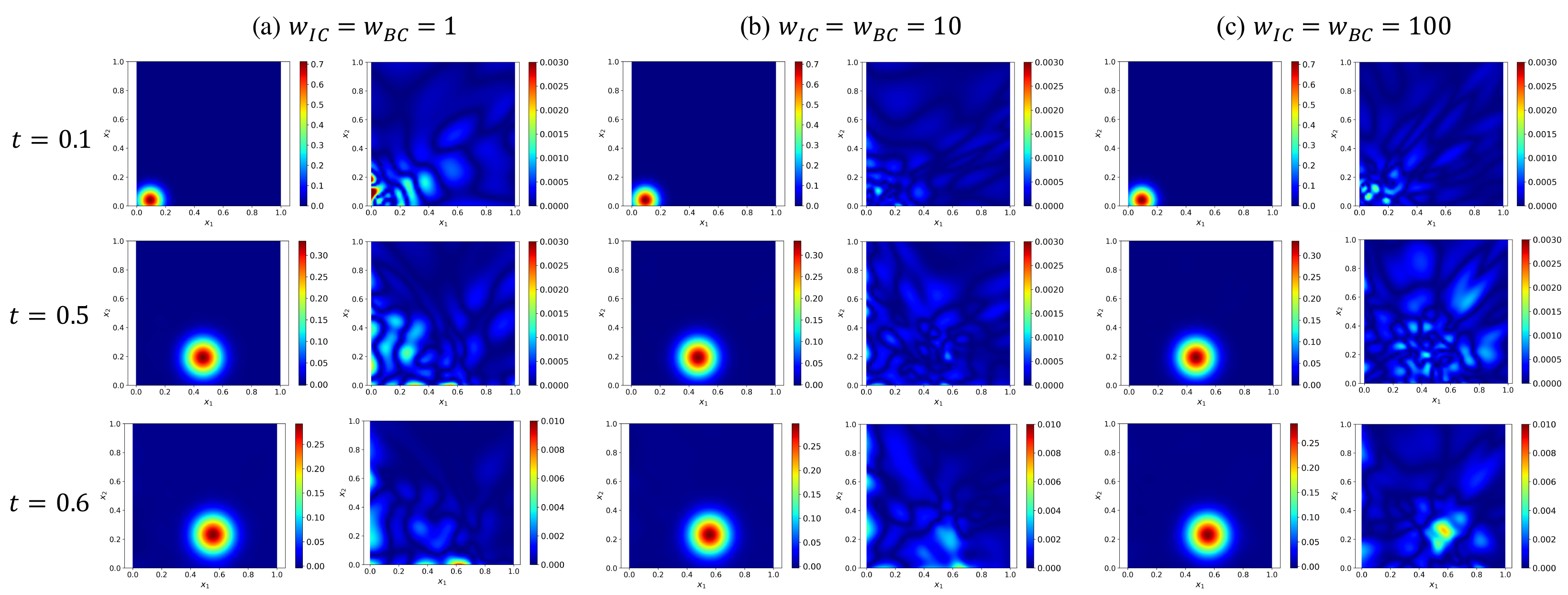}
	\caption{Same as in Figure \ref{fig:AD_2Dt_pts_Pe50}, but with $Pe=200$. For $t = 0.5$, the maximum point errors for $w=1$, $10$, and $100$ are $2.21 \times 10^{-3}$, $1.14 \times 10^{-3}$, and $1.07 \times 10^{-3}$, respectively.}
	\label{fig:AD_2Dt_pts_Pe200}
\end{figure}

% In Figures \ref{fig:AD_2Dt_pts_Pe50} and \ref{fig:AD_2Dt_pts_Pe200}, the snapshots at $t=0.6$ are the prediction obtained by PINN without using any additional data information and training procedures, where the maximun point-wise errors in these cases are nearly $1 \times 10^{-2}$. 

The snapshots of the PINN solutions at $t=0.1$, 0.5, and 0.6 for $Pe=50$ and 200 and the corresponding errors with respect to the analytical solution are shown in Figures \ref{fig:AD_2Dt_pts_Pe50} and \ref{fig:AD_2Dt_pts_Pe200}.
Here, the ADE residuals and boundary conditions are not enforced for $t>0.5$, and the PINN solutions at $t=0.6$ could be thought of as an extrapolation by the neural networks that are trained for $t \le 0.5$. 
Figures \ref{fig:AD_2Dt_pts_Pe50} and \ref{fig:AD_2Dt_pts_Pe200} show that for both $Pe$ values,  the maximum point error is less than $3 \times 10^{-3}$ for $t \le 0.5$. For $t=0.6$, the maximum errors are less than 0.01. 

% * weights
For this problem, we also investigate the effect of the $w_{BC}$ and $w_{IC}$ weights in the loss function (\ref{eq:loss_pinn}) on the accuracy of the solution ($w_f=1$ is set in all simulations). In the examples in Section \ref{sec:1D_AD}, we set $w=w_{BC} = w_{IC}= 1$. Here, we consider $w = 1$, 10, and 100.
Figures \ref{fig:AD_2Dt_pts_Pe50} and \ref{fig:AD_2Dt_pts_Pe200}  show that the weights have a significant impact on the accuracy of the PINN solution.  
For $w = 1$ (see Figures \ref{fig:AD_2Dt_pts_Pe50}a and  \ref{fig:AD_2Dt_pts_Pe200}a), the maximum errors are located near the $x_1 = 0$ and $x_2 = 0$ boundaries where the (Dirichlet) boundary conditions are prescribed. These errors are reduced as the weights increase to 10 and 100, respectively. The larger weights assigned to $J_{BC}$ and $J_{IC}$ in Eq (\ref{eq:loss_pinn}) strongly penalize the initial and boundary conditions relative to the PDE residuals that are shown to reduce the errors in the solution.
However, a very large $w$ can also lead to a decrease in accuracy because the effect of the PDE residuals will become negligible. Here, for $Pe=50$, the relative $L_2$ errors for $w=1$, $10$, and $100$ are $\epsilon = 2.41 \times 10^{-3}$, $1.49 \times 10^{-3}$, and $1.51 \times 10^{-3}$, respectively. For $Pe=200$, the relative $L_2$ errors for $w=1$, $10$, and $100$ are $\epsilon = 5.64 \times 10^{-3}$, $2.64 \times 10^{-3}$, and $4.11 \times 10^{-3}$, respectively. 
These results show that for the considered problem, $w=10$ yields optimal solutions for both $Pe$. 
The corresponding maximum point errors for $t = 0.5$ are also reported in Figures \ref{fig:AD_2Dt_pts_Pe50} and  \ref{fig:AD_2Dt_pts_Pe200}.

We note that Eq (\ref{eq:2DADE}) was solved
in \cite{Dwivedi2020} with the so-called physics informed extreme learning machine method. 
We find that the PINN method outperforms this method, as evidenced by that substantial errors that develop near boundaries in \cite{Dwivedi2020} are greatly reduced. We attribute the better performance of the PINN method to assigning larger weights to the boundary and initial condition penalty terms in the loss function.

Finally, we note that the PINN method is able to provide an accurate solution outside of the time interval where the PDE residuals and boundary conditions are enforced in the loss function. Not surprising, the accuracy of the PINN extrapolation also depends on the choice of $w$.  For example, the relative $L_2$ errors at $t=0.6$ for $Pe = 50$ are $\epsilon = 1.46 \times 10^{-2}$, $0.80 \times 10^{-2}$, and $0.91 \times 10^{-2}$ for the cases of $w=1$, $10$, and $100$, respectively.
As with the solution for $t \le 0.5$, the smallest error is achieved with  $w=10$.

%%% ---------------------------------
% 2Dt ADE-Backward
%%% ---------------------------------

\subsection{Steady-state ADE with a point source. Comparison with the finite element solutions}\label{sec:grid_effect}

\begin{figure}[htb]
	\centering
	\includegraphics[height=4cm]{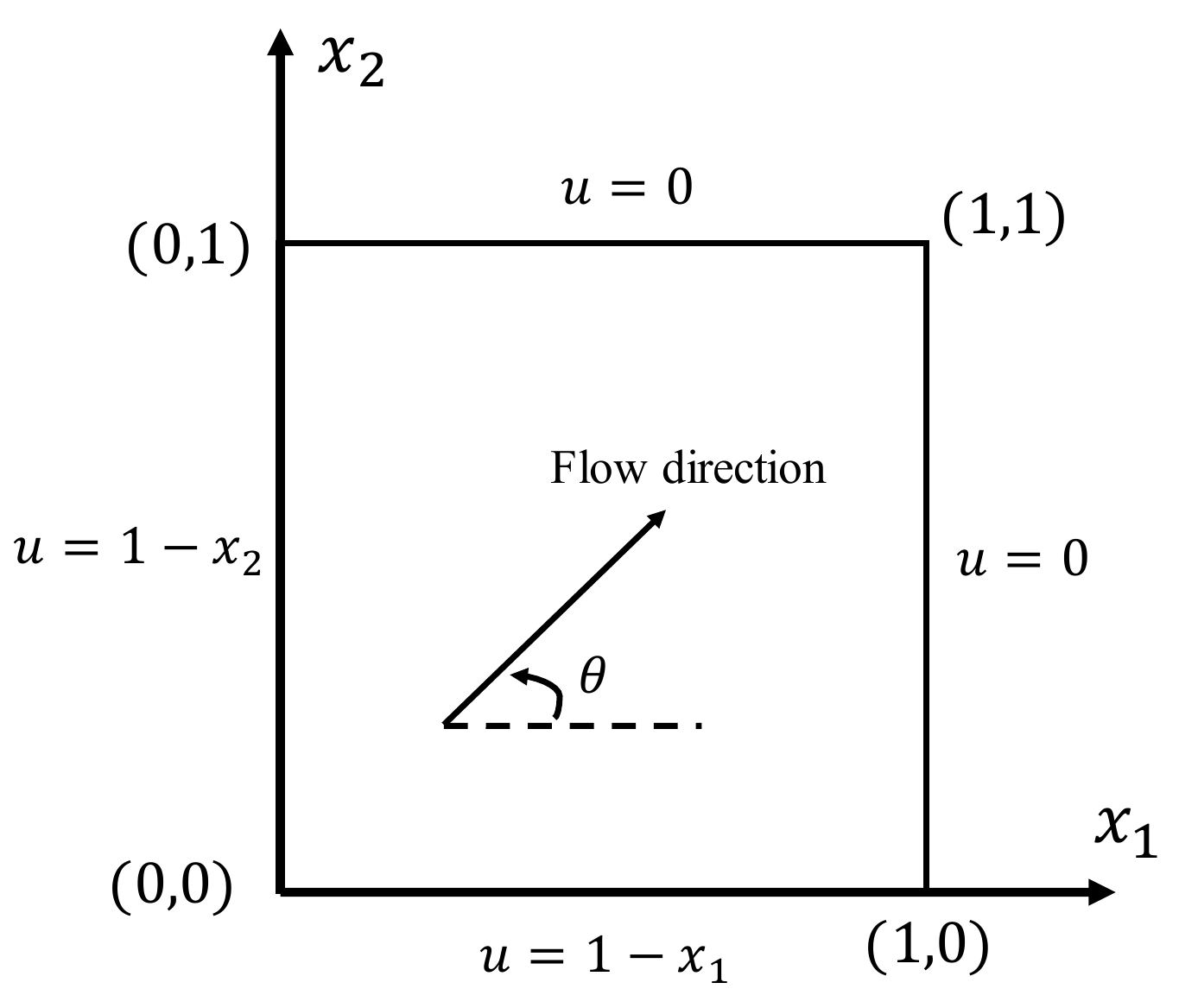}
	\caption{\small Boundary conditions for steady-state ADE with the point source located at (0,0). The velocity vector is $\vec{v}=[\cos(\pi/4), \sin(\pi/4)]^T$.}\label{fig.sche_ad_skew}
\end{figure}

\begin{figure}[htb]
	\includegraphics[height=8cm]{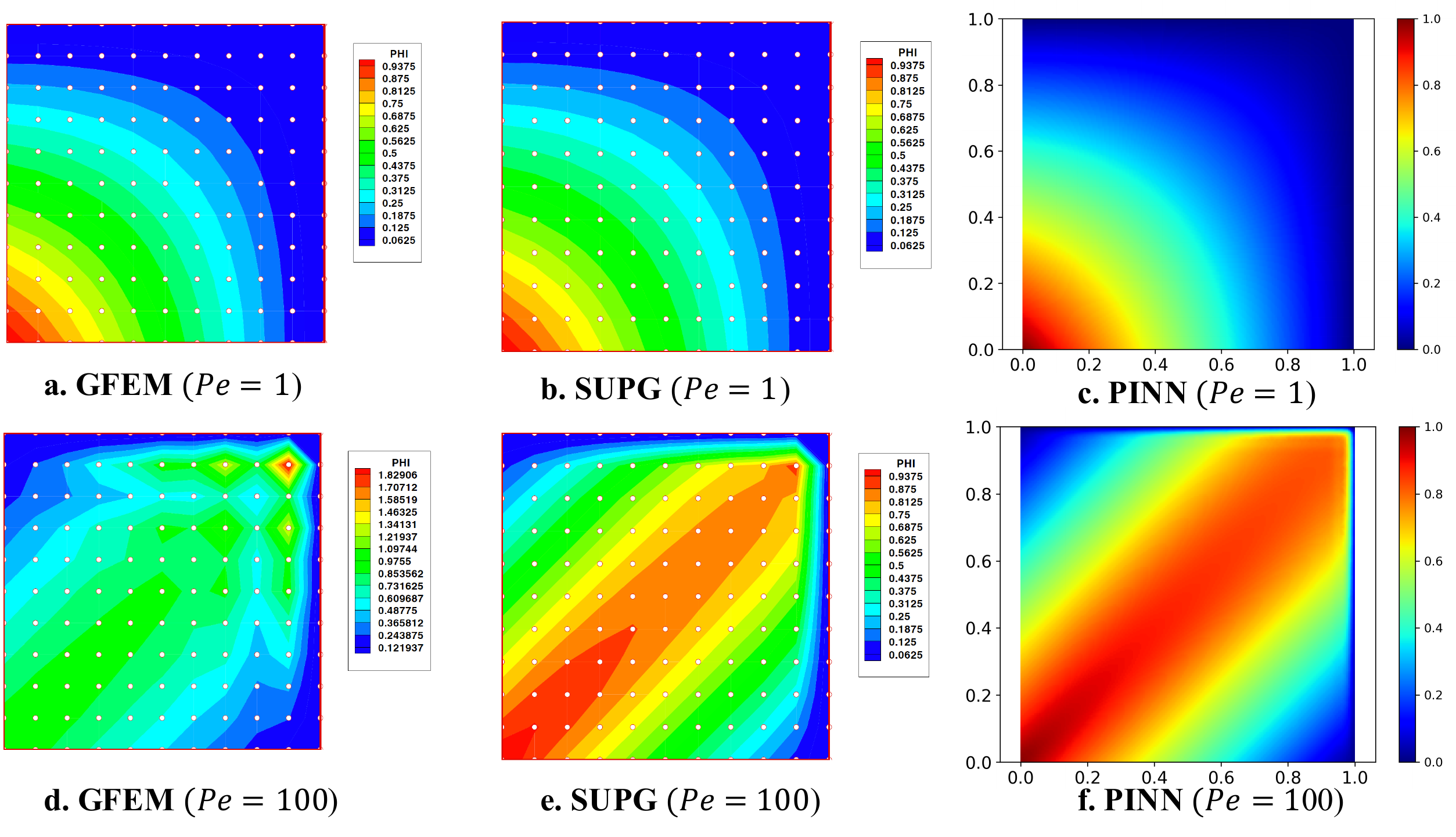}
	\caption{\small The comparison of the PINN and finite elements  solutions for the ADE with the boundary conditions and velocity vector shown in Figure \ref{fig.sche_ad_skew} for $Pe=1$ and 100. The PINN solution is obtained with $N_f = 41 \times 41$ residual points, the DNN size $3 \times 30$ and $w_{BC} = 50$.
% 	The GFEM and SUPG solutions are the reproduction from given after \cite{Lin2000}.
    The GFEM and SUPG solutions in subfigures a, b, d and e are reproduced from \cite{Lin2000}.}\label{fig.ad_skew}
\end{figure}

Here, we compare the PINN, Galerkin FEM (GFEM), and streamline-upwind Petrov-Galerkin method (SUPG) \cite{brooks1982streamline} solutions of the steady-state ADE with a point source located at $x_1=x_2=0$.  The direction of the advection velocity and the boundary conditions are shown in Figure  \ref{fig.sche_ad_skew}. The rectangular FE grid is aligned with the $(x_1,x_2)$ coordinate system and forms a $45$ degree angle with the uniform advection velocity.

The accuracy of grid-based numerical solutions depends on the grid orientation relative to the direction of flow, especially for large \textit{Pe} problems \cite{brooks1982streamline,Methods1986,Franca1992,almeida1997stable,Lin2000,Hillman2016}. Specifically, instability and/or excessive diffusion can develop if the proper crosswind diffusion is not well represented by the numerical schemes. %In this example, consider the  steady-state ADE with the advection velocity $\vec{v}=[\cos(\pi/4) \sin(\pi/4)]^T$ and the smooth Dirichlet boundary conditions as shown in Figure \ref{fig.sche_ad_skew}.

 Figures \ref{fig.ad_skew}a and b demonstrate that the GFEM and SUPG methods give similar solutions for $Pe=1$. On the other hand, for $Pe=100$, the GFEM solution develops instabilities (Figure \ref{fig.ad_skew}d),  while the SUPG solution remains stable (Figure \ref{fig.ad_skew}e) because of the use of the upwind scheme \cite{brooks1982streamline,Methods1986}. Here, the GFEM and SUPG solutions were obtained in \cite{Lin2000}.

Figures \ref{fig.ad_skew}c and f depict the PINN solutions for $Pe=1$ and 100, respectively, that are in a close agreement with the SUPG solutions. In the PINN solutions, the collocation points are uniformly spaced on the  $41 \times 41$ mesh, the boundary conditions are enforced at 41 points at each boundary, and $w_{BC} = 50$. 
While both the PINN and SUPG methods provide stable solutions for high and low $Pe$, the SUPG method requires the calibration of the stability parameter for a given mesh size and the flow direction relative to the mesh orientation.
The PINN method does not use a mesh to discretize spacial  derivatives. Therefore, the distribution of the collocation points relative to the direction of the flow does not affect the accuracy of the PINN solution.  We also observe that the solutions do not produce oscillations at the dispersion front (over- and/or undershoots) that are often present in numerical grid-based solutions of ADEs. We attribute this phenomenon to computing spatial directives analytically rather than using a numerical discretization.

It is also worth noting that it is easy to locally introduce additional residual points in the regions with high concentration gradients to further improve the accuracy of the PINN method for high $Pe$ problems. Such an approach was used in \cite{Mao2020} for fluid-flow simulations and has a similar effect to that of adaptive mesh refinement (h-refinement) in  discretization-based methods without the challenge of discretizing derivatives on a multi-resolution mesh.

%%% ---------------------------------
% 2Dt STOMP
%%% ---------------------------------
\subsection{Two-dimensional time-dependent ADE with a non-uniform velocity field}\label{sec:2DSTOMP}
So far, we have considered ADEs with a uniform velocity field and isotropic dispersion coefficient $\vec{D} = \kappa \vec{I}$, where $\vec{I}$ is the identity tensor. In this section, we solve the coupled ADE and Darcy flow equations, where the velocity field is not uniform and is given by the solution of the Darcy equation, and $\vec{D}$ is an anisotropic tensor. The ADE takes the following form:
 \begin{equation}\label{eq:ADE_heter}
 \left\{
 \begin{array}{ll}
 \begin{split}
 u_t +  \nabla \cdot [{\vec{v}} (\vec{x}) u(\vec{x})] & = \nabla \cdot [ \vec{D} \nabla u(\vec{x})], \quad \vec{x}\in\Omega,\quad t\in (0,T] \\
 u(\vec{x},t) & = u_D(x_2), \quad \quad x_1 = 0 \\
\partial u(\vec{x},t) /  \partial x_1 & = 0, \quad \quad x_1 = L_1 \\
\partial u(\vec{x},t) /  \partial x_2 & = 0, \quad \text{on} \quad x_2 = 0 \quad{\rm{and}}\quad x_2 = L_2\\
u(x,t=0) &=0, \quad \vec{x}\in \Omega
 \end{split}
 \end{array} \right.
 \end{equation}
 where $\vec{D}$ is defined in Eq. (\ref{eq:coe_dispersion})
and $\vec{v}$ is the average pore velocity
\begin{equation}\label{eq:velocity}
 \vec{v}(\vec{x})  = -\frac{K(\vec{x})}{\phi} \nabla h(\vec{x}),    
\end{equation}
and the hydraulic head $h$ is given by the steady-state Darcy equation:
 \begin{equation}\label{eq:diffusion}
 \left\{
 \begin{array}{ll}
 \begin{split}
 \nabla \cdot \left[ K(\vec{x}) \nabla h(\vec{x}) \right] & = 0, \quad \vec{x} \in \Omega \\
 h(\vec{x}) & = {H_2}, \quad {x_1} = {L_1} \\
 -K(\vec{x}) \partial h(\vec{x}) /  \partial x_1 & = q, \quad x_1 = 0 \\
 -K(\vec{x}) \partial h(\vec{x}) /  \partial x_2 & = 0, \quad x_2 = 0 \:{\rm{or}}\: x_2 = L_2.
 \end{split}
 \end{array} \right.
 \end{equation}
Here, $\phi$ is the porosity and $K(\vec{x})$ is the known hydraulic conductivity that is, in general, defined at a set of points. Usually, $K(\vec{x})$ is estimated on a mesh from a calibration study. 

In the PINN approach for Eqs (\ref{eq:ADE_heter}) -- (\ref{eq:diffusion}), we start by approximating the $K(\vec{x})$ field with a DNN $\hat{K}(\vec{x},\psi)$ that is trained using a subset of the $K$ values. The remaining values of $K$ are used to test $\hat{K}(\vec{x},\psi)$ and verify that there is no over-fitting. Next, we use PINN to solve Eq (\ref{eq:diffusion}) by approximating the hydraulic head as $h(x)\approx\hat{h}(x,\gamma)$ and computing $\gamma$ from the minimization problem

\begin{equation}\label{eq:loss_K}
\gamma = \min_\gamma \big[\frac{\omega_R}{N_f} \sum_{i = 1}^{N_f} r_f (\vec{x}_f^i,\gamma)^2  
 + \omega_{BC} (J_1 (\gamma)+J_2 (\gamma)+J_3 (\gamma)+J_4 (\gamma))
 \big], 
\end{equation}
where 
\begin{equation}
r_f (\vec{x},\gamma) =  \nabla \cdot \left[ \hat{K}(\vec{x}) \nabla \hat{h}(\vec{x},\gamma) \right] 
\end{equation}
are the residuals evaluated at $N_f$ residual points $\vec{x}_f^i$,
\begin{equation}
J_1 (\gamma) = 
\frac{1}{N_1} \sum_{i = 1}^{N_1} (\hat{h}
 (\vec{x}_{BC1}^i,\gamma) - H_2)^2
\end{equation} 
is the loss of the Dirichlet BC evaluated at $N_1$ points at $x_1=L_1$, 
\begin{equation} 
 J_2(\gamma) = 
 \frac{1}{N_2} \sum_{i = 1}^{N_2} \left(
 K(\vec{x}_{BC2}^i,\gamma)\frac{\partial \hat{h}(\vec{x},\gamma)}{\partial x_1}\big|_{\vec{x}=\vec{x}_{BC2}^i} + q \right)^2
\end{equation}
 is the loss of the Neumann boundary condition at the boundary $x_1=0$ evaluated at $N_2$ points,
 \begin{equation} 
 J_3(\gamma) = 
 \frac{1}{N_3} \sum_{i = 1}^{N_3} \left(
 K(\vec{x}_{BC3}^i,\gamma)
 \frac{\partial \hat{h}(\vec{x},\gamma)}{\partial x_2}\big|_{\vec{x}=\vec{x}_{BC3}^i} \right)^2
\end{equation}
 is the loss of the Neumann boundary condition at the boundary $x_2=0$ evaluated at $N_3$ points, and 
  \begin{equation} 
 J_4(\gamma) = 
 \frac{1}{N_4} \sum_{i = 1}^{N_4} \left(
  K(\vec{x}_{BC4}^i,\gamma)
 \frac{\partial \hat{h}(\vec{x},\gamma)}{\partial x_2}\big|_{\vec{x}=\vec{x}_{BC4}^i} \right)^2
\end{equation}
 is the loss of the Neumann boundary condition at the boundary at $x_2=L_2$ evaluated at $N_4$ points. After the DNN weights $\gamma$ are determined, the velocity is computed as
\begin{equation}\label{eq:velocity}
 \hat{\vec{v}}(\vec{x},\gamma)  = -\frac{\hat{K}(\vec{x},\psi)}{\phi} \nabla \hat{h}(\vec{x},\gamma).    
\end{equation}
 In the above, all spatial derivatives of DNNs are computed using automatic differentiation \cite{Baydin2015}. The ADE is solved as described in Section \ref{sec:method} with the residual (\ref{eq:residual_f}) written in terms of $\hat{\vec{v}}(\vec{x},\gamma)$:
\begin{equation}\label{eq:residual_f_gamma}
	r_f (\vec{z},\theta) = \hat{u}_t(\vec{z},\theta) + \nabla \cdot (-\vec{D} \nabla \hat{u} (\vec{z},\theta) + \hat{\vec{v}}(x,\gamma) \hat{u}(\vec{z},\theta)) - q, \quad  \vec{z}\in  \Omega \times (0, T). 
\end{equation}
The parameters $\gamma$ in the $\vec{v}(\vec{x},\gamma)$ DNN are frozen when training the ADE solution $\hat{u}$.

\begin{figure}[htb]
	\captionsetup[subfloat]{farskip=0.0pt,captionskip=0pt}
	\centering
	\subfloat[$K$] {\includegraphics[angle=0,width=2.3in]{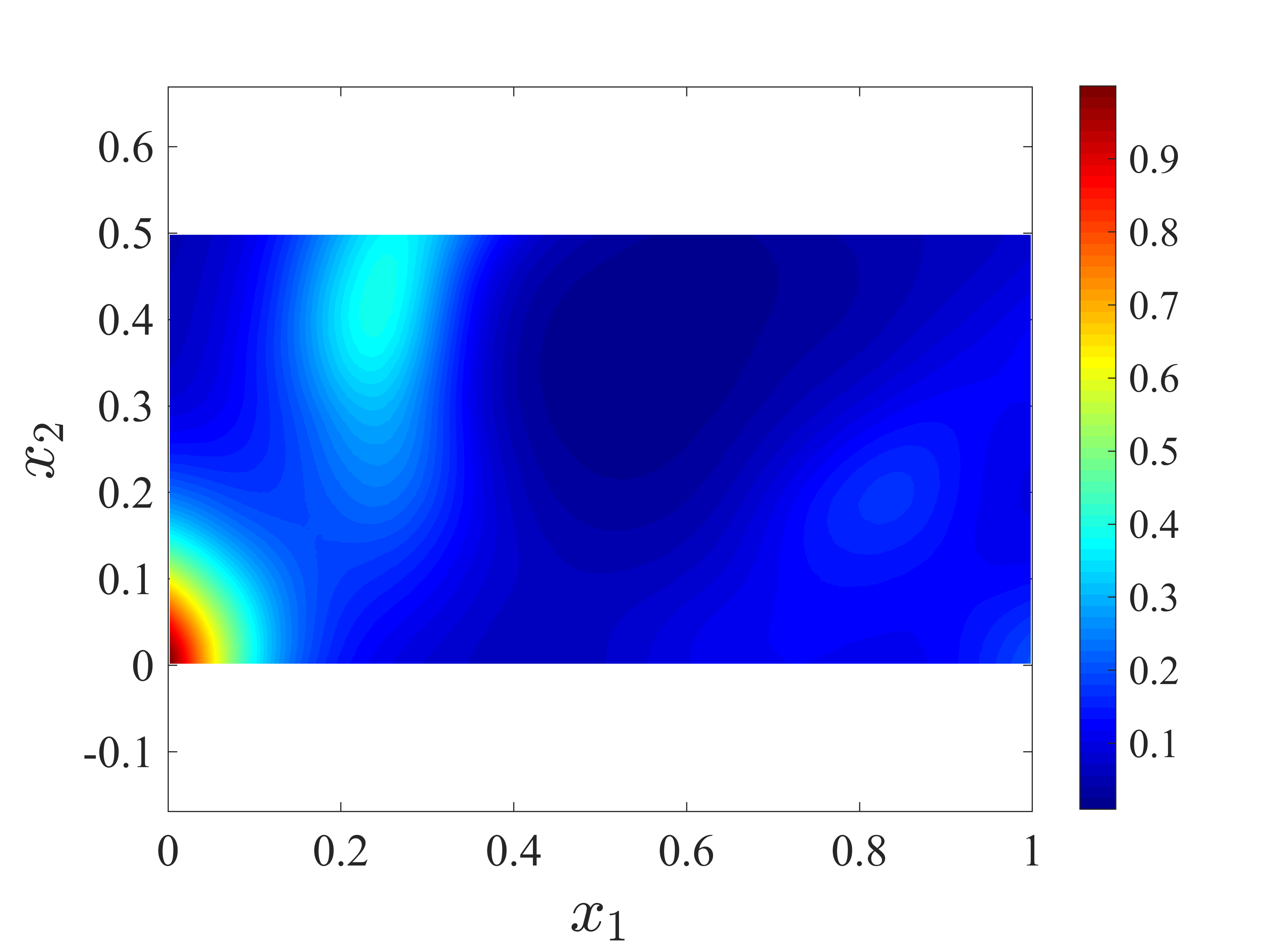}}
	%	\subfloat[$h$] {\includegraphics[angle=0,width=2in]{figures/referecen_sin_smooth/plot-head_xz.png}}
	\subfloat[$u$] {\includegraphics[angle=0,width=2.3in]{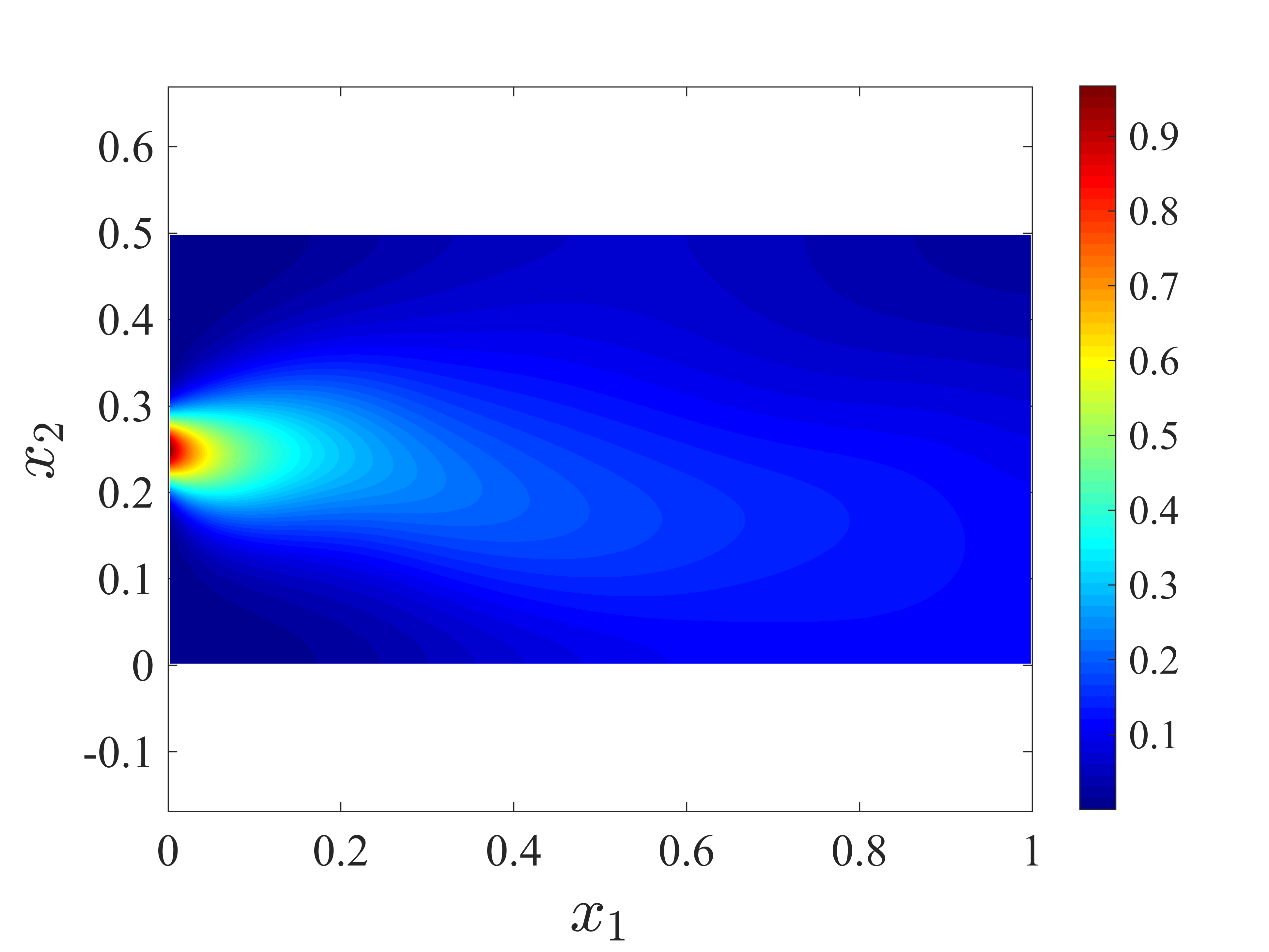}}
	\caption{\small Reference fields: (a) conductivity $K(\vec{x})$ and (b) concentration $u(\vec{x})$ at $t = 20$ min.}
	\label{fig:reference_sin_smooth}
\end{figure}

The parameters in the above equations are set to: $L_1 = 1 \: \rm{m}$, $L_2 = 0.5 \: \rm{m}$, $H_2=0 \: \rm{m}$, $q=1 \: \rm{m/hr}$, $u_D(x_2)=c\exp(-\frac{(x_2 - L_2/2)^2}{\epsilon^2})$, $c=1 \: \rm{Kg/m^3}$, $\epsilon = 0.25 \: \rm{m}$, $\phi=0.317$, $D_w=0.09 \: \rm{m^2/hr}$, $\tau = \phi^{1/3} = 0.681$, $\alpha_L = 0.01 \: \rm{m}$, and $\alpha_T = 0.001 \: \rm{m}$. 
% In this example, we consider a heterogeneous conductivity field given by $K(\vec{x}) = 0.5 \sin(4 \pi x_1)\sin(4 \pi x_2) + 1$, as shown in Figure \ref{fig:reference_sin_smooth}a. 
In this example, we consider a heterogeneous conductivity field generated as a realization of lognormal processes with the correlation length $\lambda = 0.5$, as shown in Figure \ref{fig:reference_sin_smooth}a.
Because there is no analytical solution for this problem, the reference solutions of $u(\vec{x})$ and $h(\vec{x})$ are obtained using STOMP on the $256 \times 128$ spatial grid. A snapshot of the concentration solution $u$ at $t = 20$ min is shown in Figure \ref{fig:reference_sin_smooth}b.

\begin{figure}[htb]
	\captionsetup[subfloat]{farskip=0.0pt,captionskip=0pt}
	\centering
% 	\subfloat[] {\includegraphics[angle=0,width=2.3in]{figures/AD_2Dt_CAD/map_h__r111_k600_h600_f1000_s0_pred.png}}
% 	\subfloat[] {\includegraphics[angle=0,width=2.3in]{figures/AD_2Dt_CAD/map_h__r111_k600_h600_f1000_s0_errors.png}}
	\subfloat[] {\includegraphics[angle=0,width=2.3in]{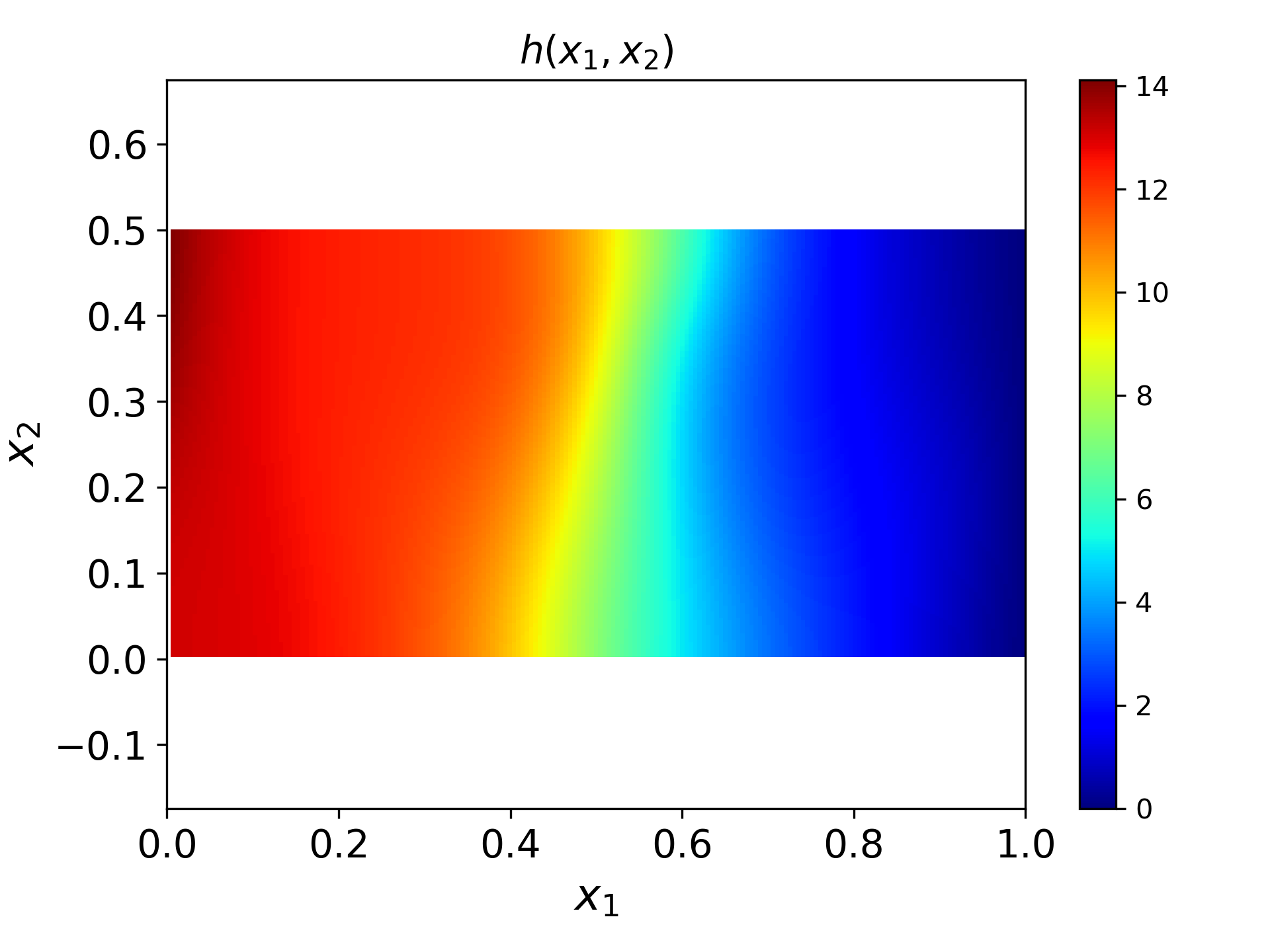}}
	\subfloat[] {\includegraphics[angle=0,width=2.3in]{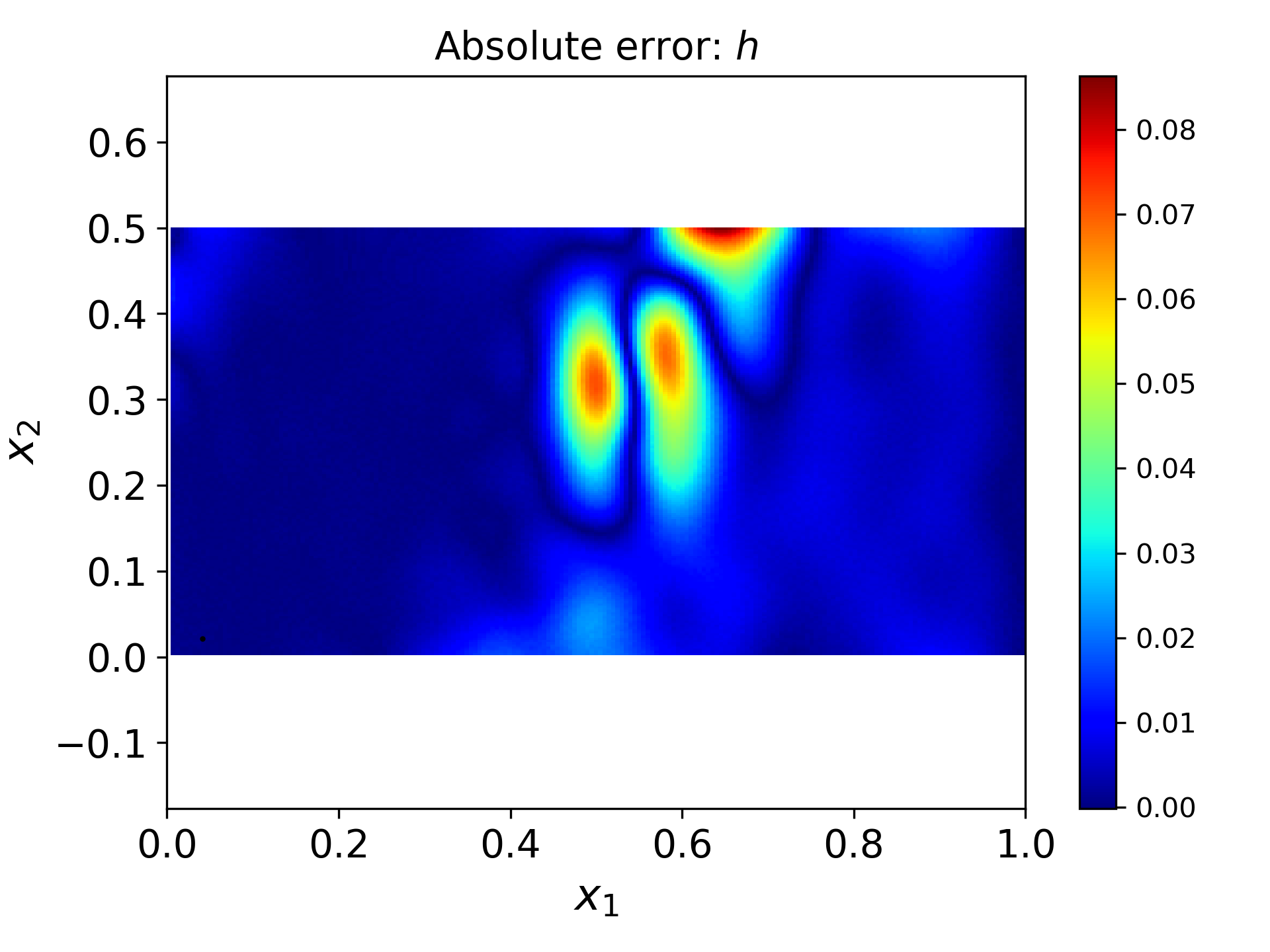}}
	\caption{\small (a) The hydraulic head $h$ obtained as a PINN solution of the Darcy equation (\ref{eq:diffusion}) and (b) the corresponding absolute point errors with respect to the reference STOMP solution.}
	\label{fig:AD_2Dt_CAD_h}
\end{figure}

\begin{figure}[htb]
	\captionsetup[subfloat]{farskip=0.0pt,captionskip=0pt}
	\centering
% 	\subfloat {\includegraphics[angle=0,width=2.3in]{figures/AD_2Dt_CAD/map_v1__r111_k600_h600_f1000_s0_pred.png}}
% 	\subfloat {\includegraphics[angle=0,width=2.3in]{figures/AD_2Dt_CAD/map_v2__r111_k600_h600_f1000_s0_pred.png}}
	\subfloat {\includegraphics[angle=0,width=2.3in]{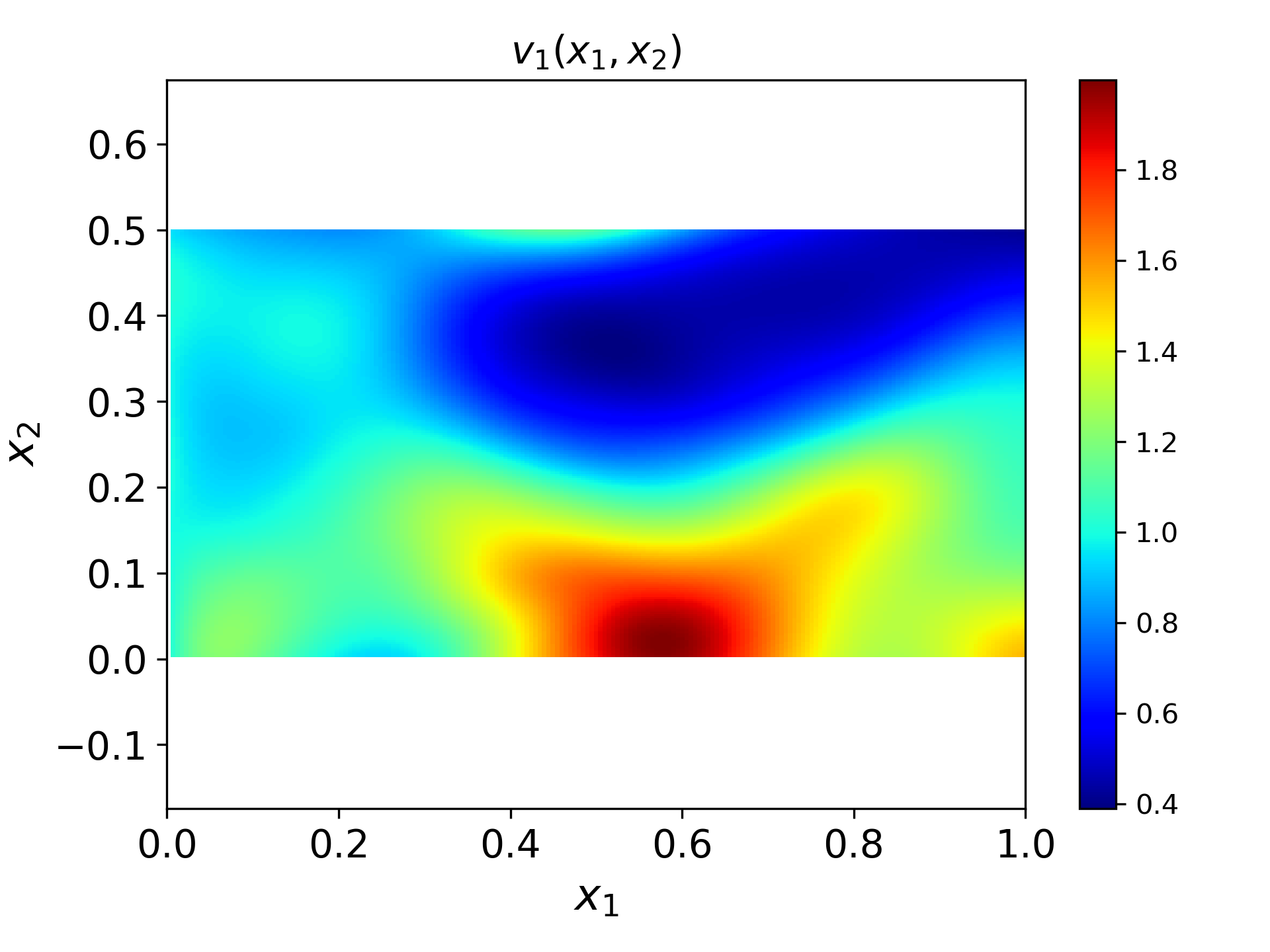}}
	\subfloat {\includegraphics[angle=0,width=2.3in]{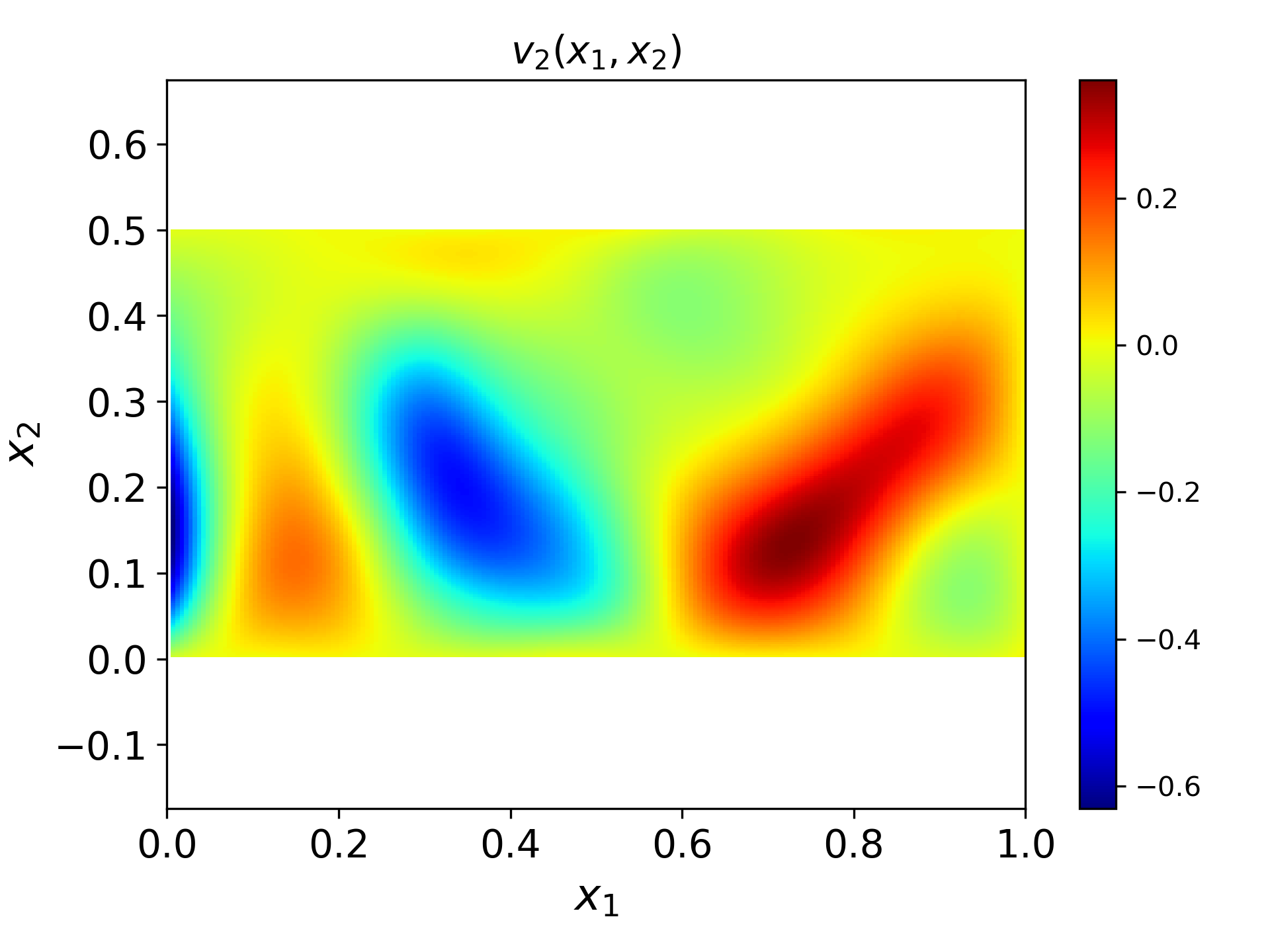}}
	\caption{\small{The $x_1$ and $x_2$ components of the velocity vector field $\vec{v}$ obtained by the PINN method from Eq. (\ref{eq:velocity}).}}
	\label{fig:AD_2Dt_CAD_velocity}
\end{figure}

% PINN for Darcy
%We first use the PINN method to solve the Darcy equation in the form of the $h(\vec{x},\theta_h)$ DNN. Then, a DNN approximation of the velocity field $\vec{v}(\vec{x},\theta_h)$ is computed from Eq. (\ref{eq:velocity}) where the gradient of the $h(\vec{x},\theta_h)$ DNN is evaluated using the
%field $\vec{v}$, the PINN approach with an independent DNN representation for $h(\vec{x})$ is first applied to solve the Darcy equation (\ref{eq:diffusion}) with the given conductivity field.
%We simply follow the same procedures of solving classical diffusion equations by PINN~\cite{Raissi2019,Tartakovsky2020a,he2020physics}. But in this example, the trained DNN $\hat{h(\vec{x},\theta_h)}$ is used to compute the velocity through automatic differentiation~\cite{Baydin2015}.
%, and the velocity derived from $\hat{h}$ will be used jointly to solve the AD equation (\ref{eq:ADE_heter}).
 
%Thus, solving the Darcy equation (\ref{eq:diffusion}) by using PINN can be viewed as a pre-training step for 
Figure \ref{fig:AD_2Dt_CAD_h} shows the PINN solution $\hat{h}(\vec{x},\gamma)$ for the hydraulic head and the point errors in this solution. We see a good agreement with the reference solution, with maximum point errors of less than 0.09 and a relative $L_2$ error of $\epsilon_h = 1.63 \times 10^{-3}$.
The velocity field given by the $\vec{v}(\vec{x},\gamma)$ DNN (\ref{eq:velocity}) is shown in Figure \ref{fig:AD_2Dt_CAD_velocity}. %Due to the heterogeneities in conductivity, we can see that the velocity is spatially-varying function \textcolor{blue}{Placeholder: what difficulties does spatially-varying velocity cause?}. 

% PINN for AD
\begin{figure}[htb]
	\captionsetup[subfloat]{farskip=0.0pt,captionskip=0pt}
	\centering
	\includegraphics[angle=0,width=5in]{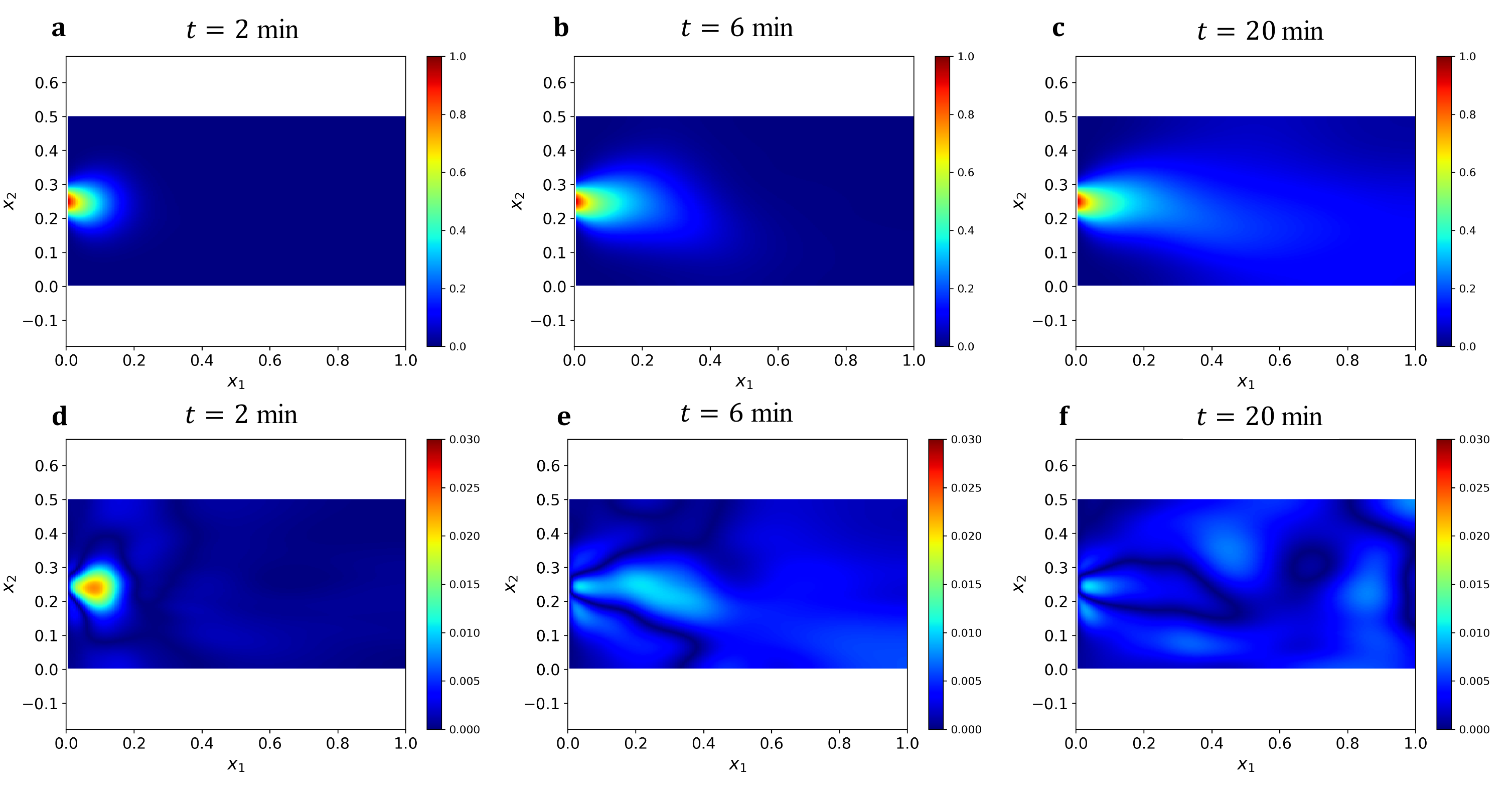}
    \caption{\small The concentration $u$ obtained as a PINN solution of ADE (\ref{eq:ADE_heter}) at (a) $t=2$ min, (b) $t=6$ min, and (c) $t=20$ min. The absolute point errors with respect to the STOMP solution are shown in subfigures (d--f). The DNN size is $5 \times 60$.} 
	\label{fig:AD_2Dt_CAD_conc}
\end{figure}

We solve the ADE (\ref{eq:ADE_heter}) for $t\in[0, 20]$ min that is enough for the solution $u(\vec{x},t)$ to reach a steady state. In the PINN method, we enforce the initial and boundary conditions on the time-space grid $40 \times 256 \times 128$, which is consistent with the time-space discretization in the STOMP simulation. The number of residual points is $N_f = 40,000$. 
In the two-step optimization algorithm, we first perform $200,000$ Adam iterations with a mini-batch size of $1000$ and the learning rate $0.0002$, followed by the L-BFGS-B optimizer. 

The PINN solution for the concentration field $u(x,t)$ and the point errors (the difference between the PINN and STOMP solutions) at $t=2$ min, $t=6$ min, and $t=20$ min are shown in Figure \ref{fig:AD_2Dt_CAD_conc}, with the relative $L_2$ errors $3.82 \times 10^{-2}$, $3.81 \times 10^{-2}$, and $2.54 \times 10^{-2}$, respectively.
% with the maximum point errors below 0.03.
% The relative $L_2$ errors  at $t = 2 \text{min}$, $t = 6 \text{min}$, $t = 20 \text{min}$ are $3.82 \times 10^{-2}$, $3.81 \times 10^{-2}$, and $2.54 \times 10^{-2}$, respectively.
The comparison of Figures \ref{fig:reference_sin_smooth}b and \ref{fig:AD_2Dt_CAD_conc}c shows a close qualitative agreement between the PINN and STOMP solutions, with the maximum point errors below 0.03. From the error plots in Figures \ref{fig:AD_2Dt_CAD_conc}d--f, we observe that the maximum errors in the PINN solution develop at the front of the plume where the concentration gradient is the largest. 
In addition, the point errors reduce with time as the plume becomes more diffused.  % We attributes this to the diffusion mechanism in the problem (\ref{eq:ADE_heter}) such that the numerical solution becomes smoother as time evolves, which is suitable for DNN approximation.

%%% ---------------------------------
% 2Dt STOMP Backward
%%% ---------------------------------

\section{Backward ADEs}\label{sec:result_back}

\subsection{Backward ADE with uniform velocity field}\label{sec:ADE_Back_Pts}
In this section, we consider the backward form of the ADE (\ref{eq:2DADE}), also known as a final boundary value problem, where the terminal condition is given at $t=T$:
\begin{equation}\label{eq:2DADE_backward}
\left\{
\begin{array}{ll}
\begin{split}
& u_t + \nabla \cdot (-\kappa \nabla u + \vec{a} u) = q, \quad \vec{x}\in \Omega =(0,1) \times(0,1) \quad t\in (0, T)\\
& \vec{a}= [\cos(\phi), \; \sin(\phi)]^T \quad \phi = 22.5^{\circ} \\
& u  = \frac{1}{4t+1}\exp{\left(-\frac{||\vec{x}-\vec{a}t||^2}{\kappa (4t +1)} \right)}, \quad \vec{x} \in  \partial \Omega\quad t\in (0, T) \\
& u (\vec{x},t=T) = \frac{1}{4T+1}\exp{\left(-\frac{||\vec{x}-\vec{a}T||^2}{\kappa (4T +1)} \right)}, \quad  \vec{x}\in \Omega
\end{split}
\end{array} \right.
\end{equation}
% \begin{equation}\label{eq:2DADE_backward}
% u (\vec{x},t=T^*) = \frac{1}{4T^*+1}\exp{\left(-\frac{||\vec{x}-\vec{a}T^*||^2}{\kappa (4T^* +1)} \right)}, \quad  \vec{x}\in \Omega
% \end{equation}
% \begin{equation}\label{eq:2DADE_backward}
% \left\{
% \begin{array}{ll}
% \begin{split}
% u_t + \nabla \cdot (-\kappa \nabla u + \vec{a} u) & = q, \quad \vec{x}\in \Omega =(0,1) \times(0,1) \quad t\in (0, T)\\
% \vec{a}= [\cos(\phi) \; \sin(\phi)]^T & \quad \phi = 22.5^{\circ} \\
% u  = \frac{1}{4t+1}\exp{\left(-\frac{||\vec{x}-\vec{a}t||^2}{\kappa (4t +1)} \right)},&  \quad  \vec{x} \in  \partial \Omega\quad t\in (0, T) \\
% u (\vec{x},t=T*) & = \frac{1}{4T^*+1}\exp{\left(-\frac{||\vec{x}-\vec{a}T^*||^2}{\kappa (4T^* +1)} \right)}, \quad  \vec{x}\in  \quad  \Omega
% \end{split}
% \end{array} \right.
% \end{equation}
The solution $u(\vec{x},t)$ is sought for $0 \leq t < T$.
Note that backward diffusion problems are normally ill-posed and that standard numerical methods require a form of regularization for solving such equations \cite{xiong2006two}.

\begin{figure}[!ht]
	\captionsetup[subfloat]{farskip=0.0pt,captionskip=0pt}
	\centering
	\includegraphics[angle=0,width=5in]{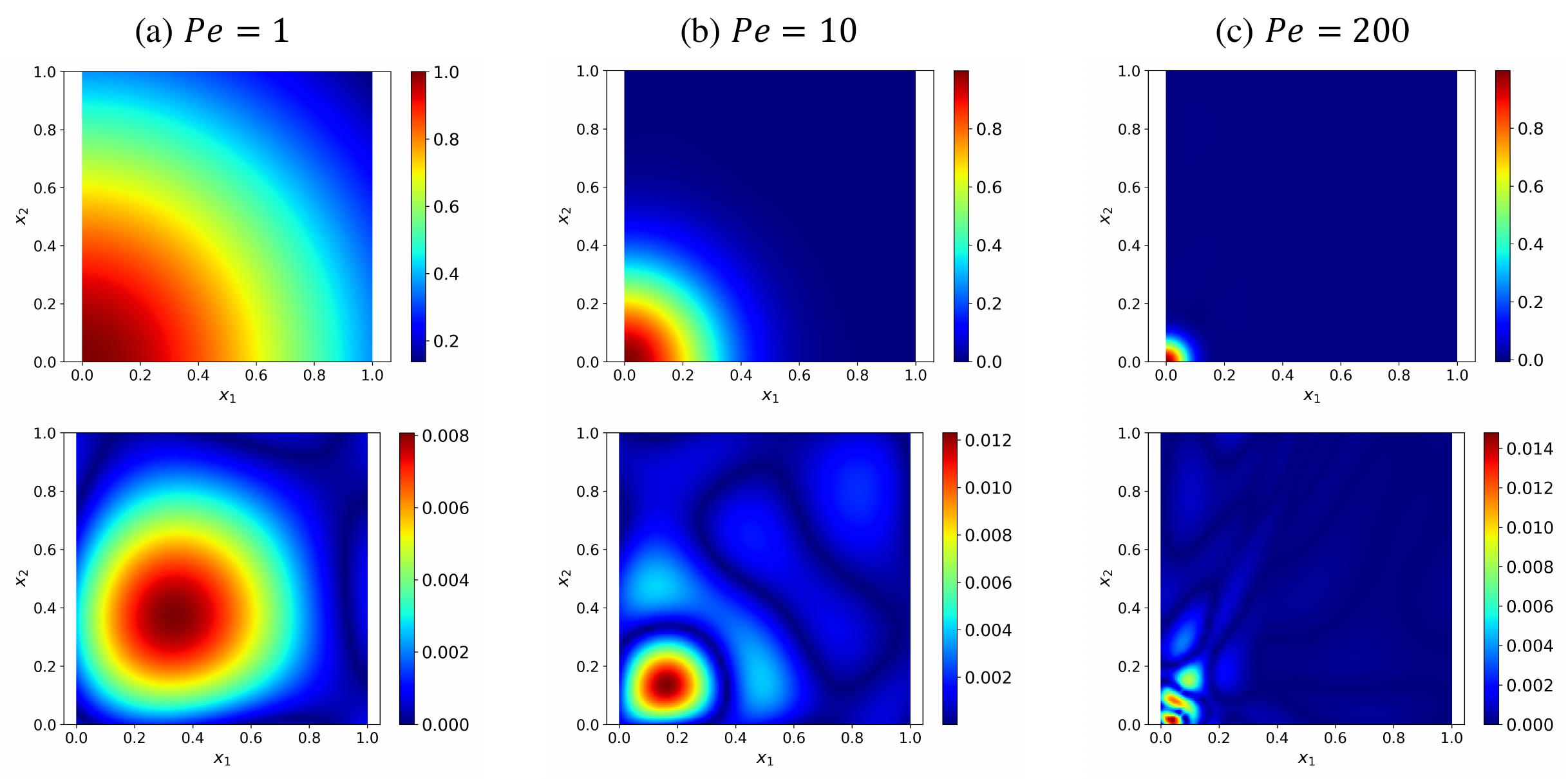}
	\caption{Upper row: the PINN solution of the backward ADE (\ref{eq:2DADE_backward}) at $t = 0$ with (a) $Pe = 1$, (b) $Pe = 10$, and (c) $Pe = 200$. Lower row: the absolute point errors with respect to the reference solutions.
	The maximum point errors for these three cases are $8.07 \times 10^{-3}$, $1.23 \times 10^{-2}$, and $1.40 \times 10^{-2}$, respectively.}
	\label{fig:AD_2Dt_pts_back}
\end{figure}

Table \ref{Table:ADE_back_Pts_err} shows the relative $L_2$ errors $\epsilon$ in the PINN solution to the backward problem (\ref{eq:2DADE_backward}) at different times ($t = 0.0, 0.1, 0.3, 0.5$) and P\'{e}clet numbers ($Pe = 1, 10, 200$).
As expected, $\epsilon$ increase as time decreases toward $t = 0$ but remains smaller than $2 \times 10^{-2}$ in all considered examples.
The PINN solutions $\hat{u}(\vec{x},t,\theta)$ at $t = 0$ and the corresponding error distributions for different \textit{Pe} are given in Figure \ref{fig:AD_2Dt_pts_back}.
The approximation errors slightly increase with $Pe$. It appears that these errors are related to the gradients in the solutions and not to the loss of information (that leads to ill-posedness of the backward ADE problem). This is important because the former source of errors can be reduced by increasing the DNN size and the number of residual points.
% This example demonstrates that the PINN approach works well for the forward and backward problems.

\begin{table}[htb]
	\centering
	\small
 	\caption{The relative $L_2$ errors of the PINN solutions of the two-dimensional backward ADE (\ref{eq:2DADE_backward}) with respect to the reference solutions for different P\'{e}clet numbers. The DNN size $3 \times 30$ and the weight $w=10$ are used in the PINN solutions.}
	%	\resizebox{\columnwidth}{!}{%}
	\begin{tabular}{cccc}
		\toprule
		Time & $Pe = 1$  & $Pe = 10$   & $Pe = 200$   \\
		\hline
% 		\multicolumn{1}{c}{$\epsilon$ in $[0,T]$}      & $2.025 \times 10^{-3}$    & $9.186 \times 10^{-3}$     & $2.019 \times 10^{-2}$   \\		
% 		\multicolumn{1}{c}{Max point err., $t = 0$}      & $8.070 \times 10^{-3}$    & $9.186 \times 10^{-3}$     & $2.019 \times 10^{-2}$   \\
		\multicolumn{1}{c}{$t = 0.0$}      & $6.205 \times 10^{-3}$    & $1.214 \times 10^{-2}$     & $2.077 \times 10^{-2}$   \\
		\multicolumn{1}{c}{$t = 0.1$}      & $8.834 \times 10^{-4}$    & $2.967 \times 10^{-3}$     & $7.344 \times 10^{-3}$   \\	
		\multicolumn{1}{c}{$t = 0.3$}      & $1.672 \times 10^{-4}$    & $7.694 \times 10^{-4}$     & $4.449 \times 10^{-3}$   \\
		\multicolumn{1}{c}{$t = 0.5$}      & $1.173 \times 10^{-4}$    & $3.318 \times 10^{-4}$     & $3.218 \times 10^{-3}$   \\
		\bottomrule
	\end{tabular}
	\label{Table:ADE_back_Pts_err}
\end{table}

\subsection{Backward ADE with non-uniform velocity field}\label{sec:2DSTOMP_backward}
% PINN for ADE-Heterogeneous
\begin{figure}[htb]
	\captionsetup[subfloat]{farskip=0.0pt,captionskip=0pt}
	\centering
	\includegraphics[angle=0,width=5in]{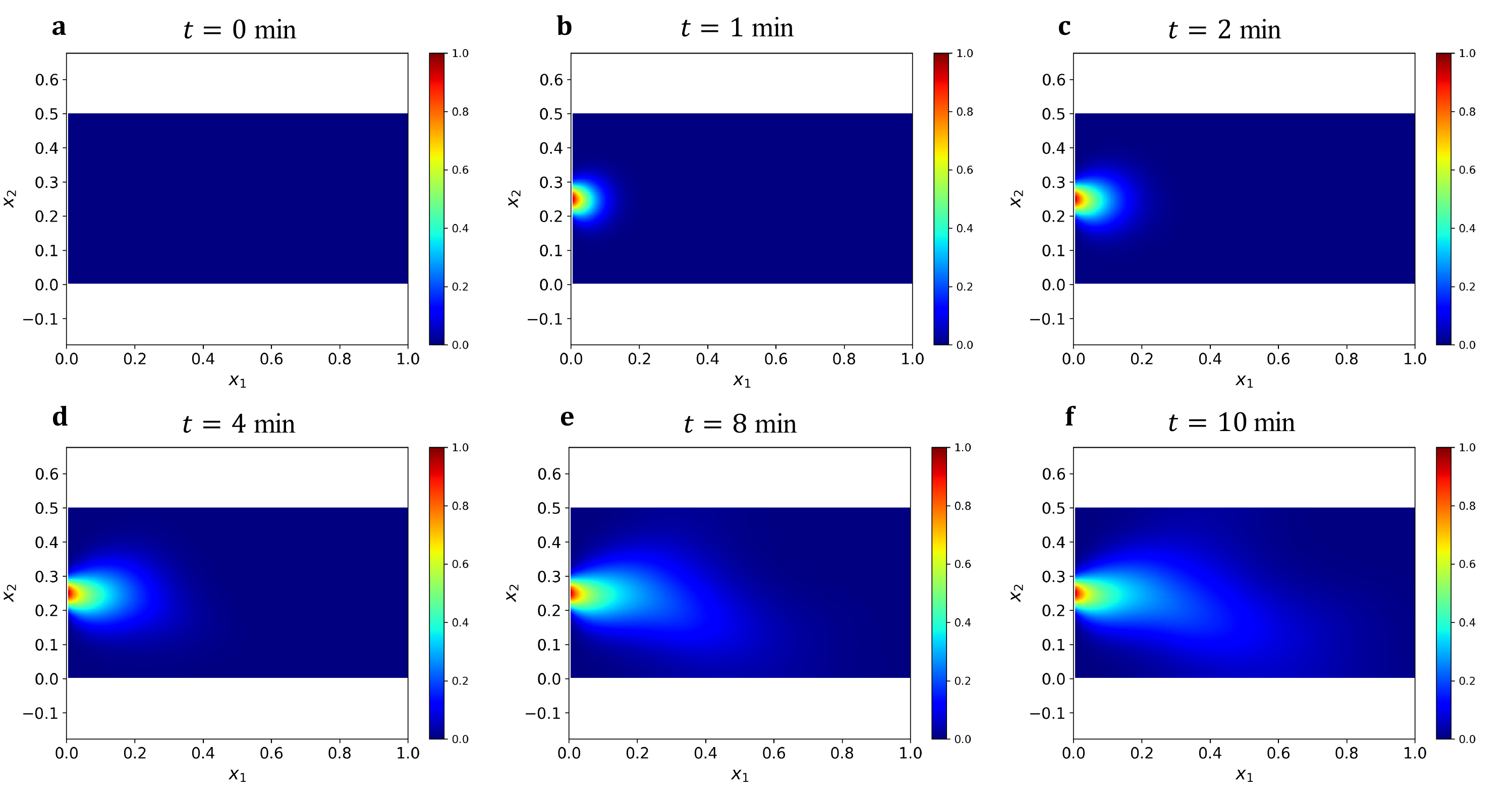}
    \caption{\small The reference concentration $u$ obtained as a STOMP solution of the ADE (\ref{eq:ADE_heter}) at different times.} 
	\label{fig:AD_2Dt_CAD_conc_1_10}
\end{figure}

Here, we consider a backward problem corresponding to the ADE in Section \ref{sec:2DSTOMP}. We assume that $u(\vec{x},t)$ at $t = T= 10$ min (see Figure \ref{fig:AD_2Dt_CAD_conc_1_10}f) is known, and the aim is to find $u(\vec{x},t)$ for $t<10$ min. The same hyperparameters as in the forward ADE problem are used (see Section \ref{sec:2DSTOMP}), except for the number of  Adam iterations is set to 400,000. We use a $21 \times 256 \times 128$ time-space grid  and  20,000 residual points.
%This number of residual points is smaller than in the forward problem because of the smaller time domain in the backward problem. 
%Our numerical tests show that using the weight $w = 1$ in the PINN loss function yields the best approximation among the cases $w = 1$, $w = 20$, and $w = 100$, and thus, only the results with $w = 1$ are presented in this study.

\begin{figure}[htb]
	\captionsetup[subfloat]{farskip=0.0pt,captionskip=0pt}
	\centering
	\includegraphics[angle=0,width=5in]{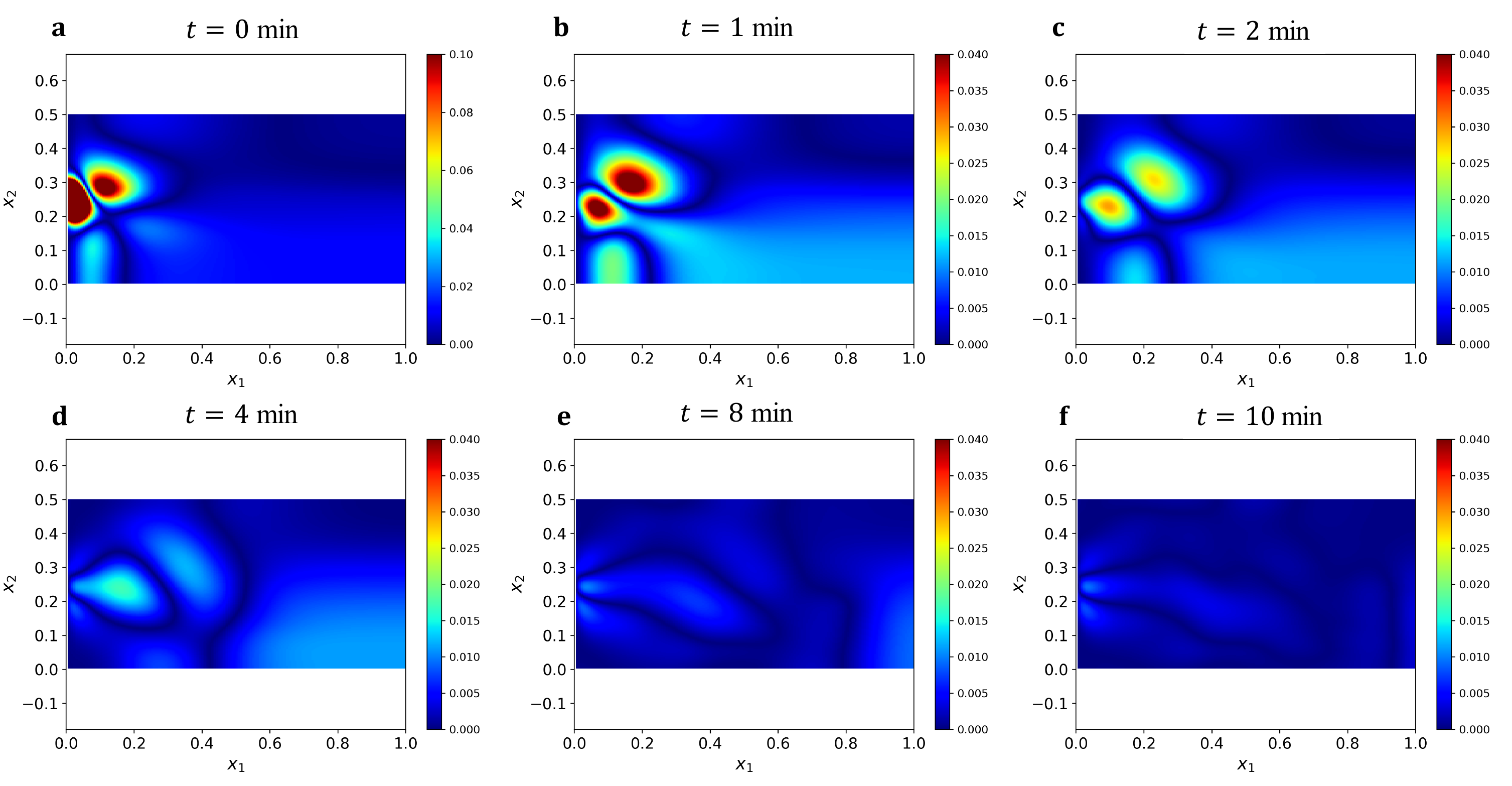}
    \caption{\small The absolute errors in the PINN solution $\hat{u}(\vec{x},t,\theta)$ with respect to the reference field $u(\vec{x},t)$ at different times. The DNN size is $5 \times 60$.} 
	\label{fig:AD_2Dt_CAD_back_err}
\end{figure}

The distributions of the absolute errors between the PINN solution and the reference concentration field $u(\vec{x},t)$ (obtained as the STOMP solution of the forward ADE in Section \ref{sec:2DSTOMP})  at $t = 0$, $1$, $2$, $4$, $8$, and $10$ min are shown in Figure \ref{fig:AD_2Dt_CAD_back_err}, and the associate maximum point errors are summarized in Table \ref{Table:ADE_back_CAD_err}. The errors increase as the backward solution evolves from $t = 10$ min to $t = 0$ min. For $t = 0$ min, a large maximum point error around 0.963 exists near the boundary $x = 0$ where the time-independent concentration $u_D$ is prescribed (refer to Eq. (\ref{eq:ADE_heter})). This is due to the discontinuity in the solution $u$ at the boundary $x = 0$ and the initial zero concentration in the reference solution.
However, for $t>0$, the maximum point errors in the backward solution do not exceed $5 \times 10^{-2}$, with the largest errors observed around the plume where large gradients develop.
% and near the $x=1$ outflow boundary where the zero normal gradient of $c$ is prescribed.
%We note that the point error near the $x=1$ boundary decreases with time as the plume moves toward this boundary and becomes more diffused.
%Also, the large errors in the backward problem appear not only around the plume but also at the region close to $x = 1$ where the plume becomes diffused, as shown in Figure \ref{fig:AD_2Dt_CAD_back_err}. This phenomenon reflects the ill-posedness of backward inverse problems for the diffusion dominant region. 

\begin{table}[htb]
	\centering
	\small
 	\caption{The maximum point errors of the PINN solution with respect to the reference STOMP solution for the backward ADE with non-uniform velocity field.}
	%	\resizebox{\columnwidth}{!}{%}
	\begin{tabular}{cccccc}
		\toprule
		$t = 0$ \text{min}  & $t = 1$ \text{min}   & $t = 2$ \text{min}   & $t = 4$ \text{min}  & $t = 8$ \text{min}   & $t = 10$ \text{min}\\
		\hline
% 		\multicolumn{1}{c}{$\epsilon_{inf}$} & 
		$9.63 \times 10^{-1}$ & $4.73 \times 10^{-2}$   & $2.97 \times 10^{-2}$ & $1.70 \times 10^{-2}$
		& $1.00 \times 10^{-2}$ & $9.34 \times 10^{-3}$ \\
		\bottomrule
	\end{tabular}
	\label{Table:ADE_back_CAD_err}
\end{table}

\section{Data assimilation}\label{sec:data_assimilation}

An important feature of the PINN method for ADEs and other PDEs is that the measurements of state variables can be incorporated in the solution without any modifications of the algorithms. When solutions are unstable (small perturbations in the solution can grow infinitely in time) or non-unique (as is the case with some backward ADEs), the state measurements can stabilize and regularize such solutions.

The measurements of $u(x,t)$, $\{ u^*(z_i) \}_{i=1}^{N_m}$ ($z_i = (\vec{x},t)_i$ is the point in time and space where the measurement is collected) can be incorporated in the PINN solution of the backward ADEs by adding the term $J_m(\theta)  = \frac{1}{N_m} \sum_{i = 1}^{N_m} (u^*(z_i) - \hat{u}(z_i,\theta))^2$ into the loss function (\ref{eq:loss_pinn}), yielding 

\begin{equation}\label{eq:loss_pinn_data}
\begin{split}
J(\theta) =  w_f J_f(\theta) + w_{BC} J_{BC}(\theta) + w_{TC} J_{TC} (\theta) + w_m J_{m} (\theta),
\end{split}
\end{equation}
%$\omega_{\gamma}, \gamma \in \{ f, BC, IC \}$
where $w_m$ is the weight corresponding to the $J_m$ loss. Note that the loss function (\ref{eq:loss_pinn_data}) for the backward equation has the term $w_{TC} J_{TC}(\theta) =w_{TC}  \frac{1}{N_{TC}} \sum_{i = 1}^{N_{TC}}  (u_{TC}(\vec{x}_i) -   \hat{u}(\vec{x}_i,t=T,\theta))^2$, which is a mean square difference with respect to the terminal condition  instead of the  $J_{IC}(\theta)$ term in the loss function (\ref{eq:loss_pinn}) for the forward ADE. We set $w_m$ and $w_{TC}$ to the same values as $w_{BC}$ in the simulations in Section  \ref{sec:2DSTOMP_backward}.

\begin{figure}[htb]
	\captionsetup[subfloat]{farskip=0.0pt,captionskip=0pt}
	\centering
	\includegraphics[angle=0,width=5in]{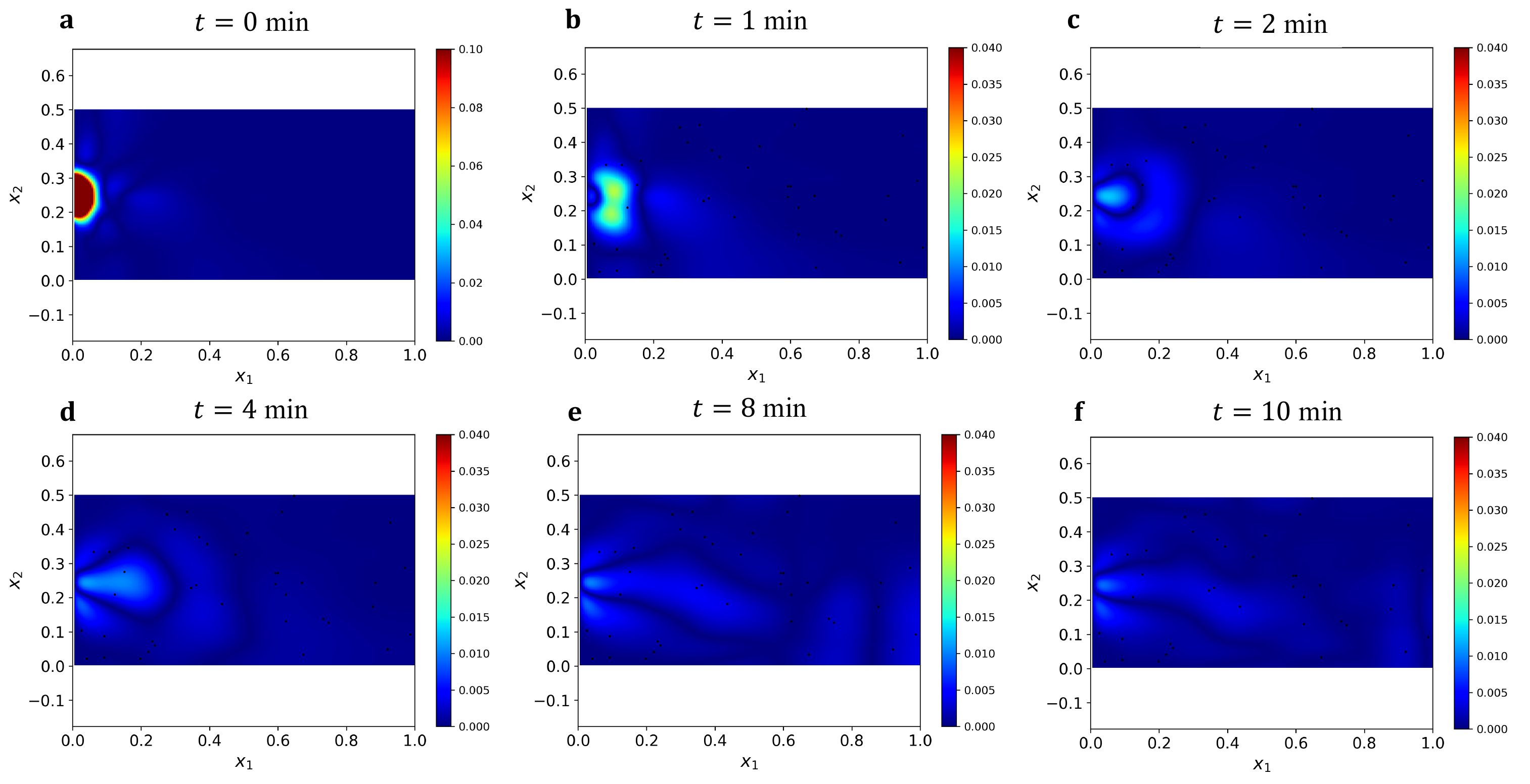}
    \caption{\small Absolute error in the PINN solution of the backward ADE with data assimilation. The absolute errors computed with respect to the reference $u(\vec{x},t)$ field. The DNN size is $5 \times 60$.} 
	\label{fig:data_assimilation}
\end{figure}

We applied the PINN method to solve the backward ADE in Section \ref{sec:2DSTOMP_backward} under the assumption that $N_m = 40$ measurements of $u$ are available at 40 spatial locations randomly distributed in $\Omega$ and uniformly distributed in time over 21 time intervals. Figure \ref{fig:data_assimilation} shows the errors in the resulting PINN solution $\hat{u}(\vec{x},t,\theta)$ with respect to the reference $u(\vec{x},t)$ field with the maximum point errors provided in Table \ref{Table:ADE_back_CAD_DA_err}. The comparison of these errors with Figure \ref{fig:AD_2Dt_CAD_back_err} and Table \ref{Table:ADE_back_CAD_err} in Section \ref{sec:2DSTOMP_backward} shows that adding 40 measurements reduces errors by more than 50\% for the prediction in $0 < t < 4 \text{min}$.   

\begin{table}[htb]
	\centering
	\small
 	\caption{The maximum point errors of the PINN solution with respect to the reference STOMP solution for the backward ADE with data assimilation.}
	%	\resizebox{\columnwidth}{!}{%}
	\begin{tabular}{cccccc}
		\toprule
		$t = 0$ \text{min}  & $t = 1$ \text{min}   & $t = 2$ \text{min}   & $t = 4$ \text{min}  & $t = 8$ \text{min}   & $t = 10$ \text{min}\\
		\hline
% 		\multicolumn{1}{c}{$\epsilon_{inf}$} & 
		$9.62 \times 10^{-1}$ & $2.29 \times 10^{-2}$   & $1.29 \times 10^{-2}$ & $1.11 \times 10^{-2}$
		& $1.00 \times 10^{-2}$ & $9.11 \times 10^{-3}$ \\
		\bottomrule
	\end{tabular}
	\label{Table:ADE_back_CAD_DA_err}
\end{table}
\section{Conclusion}\label{sec:conclusion}

%This study demonstrates that the PINN method  provide a mesh-free approach to solve challenging ADEs but with straightforward and flexible numerical implementation.Particularly, we demonstrate the effectiveness of PINN for a wide range of $Pe$, including the advection-dominated cases (associated with a large $Pe$) where sharp gradients and boundary layers could develop, which pose challenges in traditional dicretization-based methods, such as FEM and FVM. We study the effect of the weights used to penalize the residuals of the inital and boundary conditions in the loss functions, and reveal that a relative large weight values is required to ensure accurate PINN solution for forward problems. We find that the proposed PINN formulation outperforms the earlier PINN studies~\cite{Yadav2016,Khodayi-Mehr2019,Dwivedi2020}.

%For coupled Darcy flow and advection-dispersion equations with space-dependent hydraulic conductivity and velocity field, we find that the PINN solutions for the hydraulic head and concentration agree well with numerical solutions obtained with the finite volumes method (STOMP).  
%Finally, we show that the PINN method provides accurate solutions for the backward ADEs in both uniform and non-uniform advection velocity fields.

We present the PINN method for solving the coupled Darcy equation and ADE and test it for one- and two-dimensional forward and backward ADEs for a range of $Pe$. We use a weighted sum of residual terms in the loss function and show that the residuals of the initial and boundary conditions should be weighted larger than the PDE residuals to obtain accurate solutions. 
We find that the proposed PINN formulation  outperforms the earlier PINN studies~\cite{Yadav2016,Khodayi-Mehr2019,Dwivedi2020}.
For coupled Darcy flow and advection-dispersion equations with space-dependent hydraulic conductivity and velocity fields, we find that the PINN solutions for the hydraulic head and  concentration agree well with numerical solutions obtained with the finite volumes method.  
We also show that for advection-dominated transport, the PINN method outperforms the GFEM and is comparable in accuracy with the SUPG methods, the latter method requiring the calibration of a stability parameter specific to the given meshes and the flow direction.  Next, we demonstrate that the PINN method remains accurate for the backward ADEs, with the relative errors in most cases staying under 5\% compared to the reference concentration field. Finally, we show that when available, the concentration measurements can be easily incorporated in the PINN method and significantly improve (by more than 50\% in the considered cases) the accuracy of the PINN solution of the backward ADE.

% ===============================================================================
%%%%%%%%%%%%%%%%%%%%%%%%%%%%%%%%%%%%%%%%%%%%%%%%%%%%
%%%     End of body of article     %%%
%%%%%%%%%%%%%%%%%%%%%%%%%%%%%%%%%%%%%%%%%%%%%%%%%%%%
\section*{Acknowledgements}
This research was partially supported by the U.S. Department of Energy (DOE) Advanced Scientific Computing (ASCR) program. Pacific Northwest National Laboratory is operated by Battelle for the DOE under Contract DE-AC05-76RL01830. The data and codes used in this paper are available at  https://github.com/qzhe-mechanics/Repo-PINN-ADE.git.

%\bibliography{reference_PINN_AD.bib,mybib_ADE,mybib_PINN.bib,mybib_PINN_other}
\bibliography{mybib_ADE_WRR.bib}

% ===============================================================================

\end{document}